\documentclass[aps,pre,showpacs,color,xcolor,longbibliography]{revtex4-2}

\usepackage{hyperref}
\usepackage{subcaption}  
\usepackage{soul}
\setstcolor{lightgray}
\usepackage{mathtools}
\mathtoolsset{showonlyrefs}
\usepackage[normalem]{ulem} 

\usepackage{caption}

\usepackage{amsfonts}
\usepackage[T1]{fontenc}
\usepackage[latin9]{inputenc}
\usepackage{geometry}
\usepackage{color}
\usepackage{float}
\usepackage{amsmath}
\usepackage{amssymb}
\usepackage{graphicx}

\usepackage{comment}
\newcommand{\EatOneArg}[1]{}

\def\sign{\mathrm{sign}}

\def\hat{\widehat}

\def\R{{\mathbb R}}

\def\R{{\bf R}}

\def\Z{{\mathbb Z}}

\def\R{\mathbb{R}}

\def\T{{\mathbb T}}

\newtheorem{theorem}{Theorem}[section]

\newtheorem{definition}[theorem]{Definition}
\newtheorem{remark}[theorem]{Remark}

\definecolor{bluegreen}{rgb}{0.0, 0.3, 0.8}
\definecolor{darkgreen}{rgb}{0.0, 0.7, 0.5}

\definecolor{green}{rgb}{0.0, 0.5, 0.0}

\newcommand{\cfp}[1]{\color{bluegreen} {\tt [FP: #1]} \color{black}}
\newcommand{\fp}[1]{\color{bluegreen} { #1} \color{black}  }

\usepackage{xcolor}

\begin{document}
\title{Space-time resonances in the spatiotemporal spectrum of nonlinear dispersive waves}

\author{Michal Shavit$^1$, Fabio Pusateri$^2$, Zhou Zhang$^3$, Yulin Pan$^3$, Davide Maestrini$^{4}$, Miguel Onorato$^{4,5}$ and Jalal Shatah$^1$}
\affiliation{$^1$ Courant Institute of Mathematical Sciences, New York University, NY 10012, USA}
\affiliation{$^2$
Department of Mathematics, University of Toronto, Toronto M5S 2E4, Ontario, Canada}
\affiliation{$^3$ Department of Naval Architecture and Marine Engineering, University of Michigan, Ann Arbor, MI 48109, USA}
\affiliation{$^4$ Dipartimento di Fisica, Universit\`a degli Studi di Torino, 10125 Torino, Italy}
\affiliation{$^5$ Istituto Nazionale di Fisica Nucleare, INFN, Sezione di Torino, 10125 Torino, Italy}

\date{\today}

\begin{abstract}
In weakly nonlinear dispersive wave systems, long-time dynamics are typically governed by time resonances, where wave phases evolve coherently due to exact frequency matching. Recent advances in spatio-temporal spectrum measurements, however, reveal prominent features that go beyond the predictions of time resonance theory. In this work, we develop a theoretical framework to interpret these signatures by identifying and characterizing an alternative mechanism: space resonances. These arise when wave packets share the same group velocity and remain co-located, leading to long-lived interactions.  We further show that gauge-breaking terms in the Hamiltonian give rise to space resonances supported on negative frequencies. 
By combining sea-surface elevation data, numerical simulations, and analytical theory, we derive the leading-order spatio-temporal spectrum for weakly interacting water waves, providing a unified explanation for its observed features.
\end{abstract}

\maketitle

\tableofcontents

\medskip
\section{Introduction}
For non-linearly interacting waves, a central question is to determine which interactions govern the evolution at different time scales. 
A natural lens for addressing this question is the spatio-temporal Fourier transform which reveals how energy is distributed in wavenumber and frequency.
\smallskip

In the case of small amplitude dispersive waves, long-time behavior is often attributed to \emph{time-resonant interactions} between Fourier modes. $N$ waves of the form $A(k_\ell)e^{i\omega(k_\ell)t}$, $\ell = 1, \dots, N$, where $\omega(k)$ denotes the linear dispersion relation, are said to be \emph{time-resonant} if:
\begin{equation}\label{introTR}
{k}_1 \pm {k}_2 \pm \cdots \pm {k}_N = 0,
\qquad
\omega({k}_1) \pm \omega({k}_2) \pm \cdots \pm \omega({k}_N) = 0,
\end{equation}
for some combination of signs. When these conditions hold, oscillatory phases cancel, interactions do not average out, and energy may transfer persistently among modes.

\smallskip

\begin{figure}[h!]
\centering
\includegraphics[scale=0.3]
{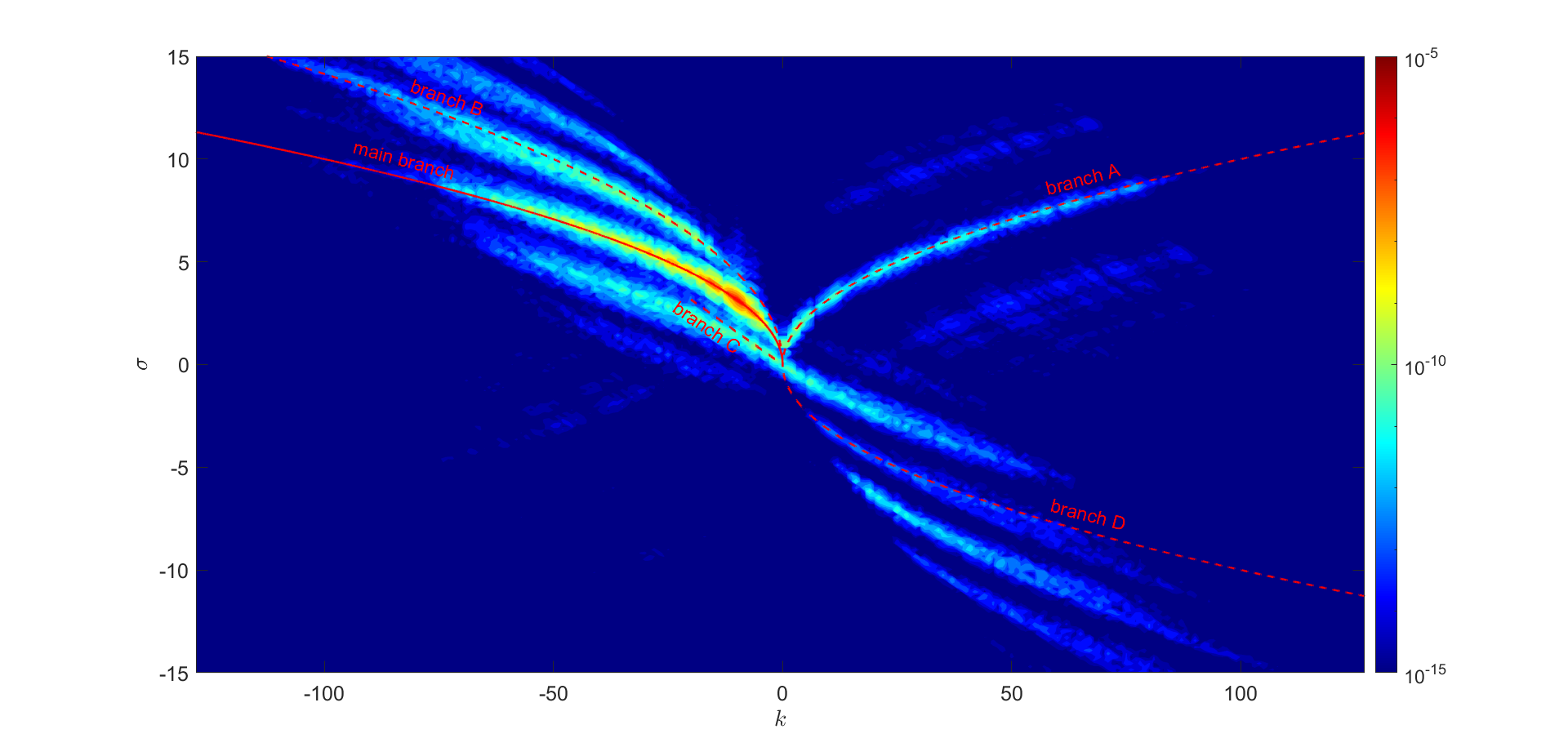} 
\caption{\label{fig:k-w_simwater}  Spatio-temporal spectrum of a linear combination of surface elevation and velocity potential on the surface, $|a(k,\sigma)|^2$, from the numerical simulation of water wave equations. Quantities in the figure are non-dimensional, corresponding to rescaled quantities in the simulation. Five branches discussed in the text are marked in the figure, including the main branch (fitted by $\sigma=\sqrt{g|k|}$ for negative $k$), branch A ($\sigma=\sqrt{g|k|}$ for positive $k$), branch B ($\sigma=\sqrt{2g|k|}$ for negative $k$), branch C (approximately by $\sigma = c_g k$ near the origin) and branch D ($\sigma=-\sqrt{g|k|}$ for positive $k$).}
\end{figure}

To further analyze this problem, it is now possible, due to recent experimental and numerical advances, to measure wave fields simultaneously in space and time, enabling the computation of the \emph{spatio--temporal spectrum}: the wavenumber--frequency (denoted by $(k,\sigma)$) spectrum via a Fourier transform in both variables. These data reveal, however, that the Fourier transform of a small-amplitude water-wave surface $\eta(x,t)$ is \emph{not} spread uniformly, nor confined to the linear dispersion curve alone. Instead, energy localizes sharply along additional surfaces in $(k,\sigma)$ that cannot be predicted from time-resonance conditions alone, see Figure \ref{fig:k-w_simwater} and \cite{leduque2025deep, falcon2022experiments,zhang2022numerical, campagne2019identifying, campagne2018impact, aubourg2017three, herbert2010observation} . Identifying the mechanism that selects these localization surfaces is therefore essential for understanding which nonlinear interactions actually dominate the dynamics.


Surface gravity waves represent one of the most celebrated and enduring problems in fluid mechanics \cite{craik2004origins}. Their equations are remarkably rich, capturing phenomena from gentle ripples to tsunamis, yet notoriously difficult to analyze due to their full nonlinearity and nonlocality. The subject has a long distinguished history, spanning oceanography (Hasselmann's kinetic theory \cite{hasselmann1962non}), mathematical physics (Zakharov's Hamiltonian formulation \cite{zakharov1966energy}), and rigorous analysis (from Stokes' expansions \cite{stokes1847theory} to modern results on well-posedness and asymptotics see \cite{lannes2013water, ionescu2018recent} and references there-in). Despite this maturity, the mechanism responsible for selective amplification of some non-resonant branches in the case of many interacting waves remained unidentified.

\smallskip
In this manuscripts we study analytically the spatiotemporal spectrum and reveal that this localized structure arises from a second class of long-lived interactions, which we call \textit{space-resonances}. Two interacting wave packets may exchange energy coherently if they share the same group velocity, without being time-resonant. Such packets remain co-located in space, creating a sustained interaction that leaves a clear spectral signature. These interactions concentrate energy along non-linear branches in $(k,\sigma)$, away from the main dispersion curve where time-resonances reside. Space-resonances thus explain the strongly enhanced, selective appearance of certain non-resonant harmonic surfaces observed in the spectrum. 

While gravity waves are our central example, our framework extends naturally to nonlinear optics, plasma waves, Bose-Einstein condensates, and other dispersive systems. We first develop the theory in a simplified model of nonlinear wave interaction, obtain an analytic leading-order formula for the spatio-temporal spectrum, then apply it to water-wave data and verify its predictions. We present numerical explorations of other systems as well.

\smallskip
In nonlinear systems, the spectrum distinguishes \emph{free modes}, which are waves 
that share the same signature as the linear dispersion relation, from \emph{bound modes}, which arise through nonlinear interactions and deviate from the linear dispersion relation. These two classes of waves produce distinct signatures in $(k,\sigma)$--space. 
A commonly adopted approach for analyzing these modes is to apply perturbation theory and normal form transformations to eliminate non-resonant terms from the system~\cite{zakharov1999statistical,krasitskii1994,janssen2009some}. 
One then obtains new variables representing the free wave field, which oscillate at the linear frequency (with nonlinear corrections) and whose dynamics are governed by resonant and near-resonant interactions, described within wave turbulence theory by the Wave Kinetic Equation. 

\smallskip

In \cite{zakharov1999statistical}, Zakharov analyzed surface gravity waves using the perturbative and normal form approach described above. For deep water, exact three-wave time resonances are excluded by the concave nature of the dispersion relation, so a normal form transformation is applied to obtain the dynamics of the free wave field. From this framework, the spectral wave action density associated with the physical water wave variables (surface elevation and velocity potential) can be analytically reconstructed starting from the spectrum of the free waves \cite{janssen2009some}. 
Although the bound modes themselves are not resonant, their inclusion in the normal form transformation modifies the statistical properties of the wave field, influencing many statistical observables \cite{janssen2009some}. 
In this procedure, all non-resonant modes are treated on equal footing and are therefore expected to appear uniformly in the spatio--temporal spectrum. However, subsequent experimental studies revealed that some non-resonant modes are more strongly enhanced than others, as demonstrated in Figure \ref{fig:k-w_simwater}. For single nearly monochromatic waves this can be attributed to Stokes harmonics, but in broadband multi-wave systems this explanation fails. The selective amplification seen in real spectra therefore requires a different mechanism.

\smallskip
We show that the missing mechanism is space-resonance, and that even when higher-order time-resonances arise, the spectral imprint of space-resonances persists.
We argue that, in the absence of time resonances, the dynamics are governed by space-resonant interactions.

The space resonance was originally introduced in a purely mathematical context, where it played a central role in proving long-time existence and scattering results for dispersive PDEs. In particular, the absence of space-resonances was used to improve long-time existence for small-amplitude water-wave solutions \cite{GMS2,GMSC,DIPP,ionescu2015global}, while their intersection with time-resonances was used to show how linear scattering is modified in the integrable 1D Schrodinger equation \cite{kato2011new} and 3D wave equations \cite{deng2020global}. Their presence without time-resonances, however, has not been investigated previously. Here we study precisely this regime, focusing on leading-order nonlinear interactions in which time-resonances are absent. We demonstrate that space-resonances redistribute energy across Fourier space and thereby shape the evolving solution, dominating the dynamics over intermediate nonlinear times. 



To briefly illustrate the concept of space-resonances and their relation to the spatio-temporal spectrum, 
let us consider the interaction of two wave packets in a weakly nonlinear three-wave system. 
A \emph{wave packet}, i.e., a localized superposition of plane waves with wavenumbers near ${k}_0$, has the form
\[
\psi({x},t) = \int_{\R^d} A({k}-{k}_0)\, e^{i({k} \cdot {x} - \omega({k})t)}\, d{k},
\]
where $A$ is 
strongly localized. 
The envelope propagates at the \emph{group velocity}
\[
v_g := \nabla_k \omega({k}_0),
\]
which represents the velocity at which  energy or information are transported.
When two such packets interact, they produce a new packet with Fourier transform
\[
I(t,{k})= \int_{\mathbb R^{d}} 
  B( k, {k}_1) 
  \, e^{it \phi({k},{k}_1)} \, 
  d{k}_1, \qquad \phi({k},{k}_1) := 
  \omega({k}_1) + \omega({k}-{k}_1),
\]
where $B$ contains amplitude factors. 
Applying the stationary phase lemma gives, for $t\gg 1$,
\[
I(t,{k}) = B( {k},{k}_1^*)
  \, e^{i t \phi({k},{k}_1^*)} \left( \frac{2\pi}{t} \right)^{d/2}
\frac{e^{i\frac{\pi}{4} \, 
\operatorname{sgn}(\nabla^2_{{k}_1} \phi({k},{k}_1^*))}}{
  \sqrt{|\det \nabla^2_{{k}_1} \phi({k},{k}_1^*)|}} + \mathcal{O}\!\left( t^{-(d+1)/2} \right),
\]
where ${k}_1^*={k}_1^*({k})$ satisfies
\begin{align}\label{introSR}
\nabla_{{k}_1} \phi( k, {k}_1^*) = 0.
\end{align}
The solutions of \eqref{introSR} 
correspond to interacting packets with identical group velocities, i.e., {\it space-resonant} wave numbers. 
In the $({k},\sigma)$ spatio-temporal spectrum their contributions are shown to be strongly localized near
\begin{align}\label{introSRsigmak}
\sigma = \phi({k},{k}_1^*).
\end{align}
Notice how \eqref{introSRsigmak} always includes the curve $\sigma = 2\omega({k}/2)$; 
this curves is evident in numerical and experimental data, such as those provided in Section \ref{sec:expnum} and Figure \ref{fig:k-w_simwater}.
Several other features of the spatio-temporal spectrum of weakly interacting dispersive systems 
can also be explained by the above theoretical framework,
including the appearance of branches with negative $\sigma$ (for systems that break gauge invariance),
and the presence of an almost 
straight branch near zero modes.



\medskip
{\bf Organization of the paper.}
The paper is organized as follows.  
Section \ref{sec:expnum} presents spatio-temporal spectra from experiments
on surface gravity waves and from numerical models with quadratic nonlinearities 
that generate non-resonant three-wave interactions.
In addition to the known second- and higher-order branches, 
we identify a branch near the zero mode and observe negative-frequency components~\cite{zaleski2020anomalous, villois2025anomalous}.  
Section~\ref{sec:space_res} develops a simplified theoretical model for space resonances and 
Section \ref{sec:waterwaves}  
applies it to the weakly nonlinear water-wave equations to obtain the leading-order spatio-temporal spectrum.

\medskip
\section{The spatiotemporal spectrum from experimental and numerical data} 
\label{sec:expnum}
Before tackling the problem from a theoretical perspective, we find it useful for the clarity of the paper to present first some experimental and numerical approaches in which we compute the spatio-temporal spectrum. 
\subsection{Stereoscopic measurements of surface gravity waves from the Acqua Alta Tower} \label{sec:expwater}
Historically, surface gravity wave measurements have been carried out using single-point instruments, such as buoys or wave gauges, which record time series data from which classical frequency spectra can be extracted. More recently, stereoscopic techniques have made it possible to capture both spatial and temporal information over areas spanning about thousand square meters \cite{benetazzo2006measurements,benetazzo2012offshore}. The dataset examined in this study was obtained from the Acqua Alta oceanographic research tower, located in the northern Adriatic Sea (Italy) at 15 km offshore from the Venetian coast, in 17 meters of water. Data were acquired using a pair of cameras simultaneously capturing the same area. Sea surface elevation, $\eta(x,y,t)$, was measured using a stereographic method \cite{benetazzo2006measurements,bergamasco2017wass}. The resulting dataset consists of approximately 28 minutes of recordings over a square area with side length $\ell = 32.6$ m, sampled at a frequency of 12 Hz. More information on the data set can be found in \cite{guimaraes2020data}. 
The dataset represents a suitable framework for studying the spatio-temporal spectrum. 
\begin{figure}[h!]
\centering
\includegraphics[scale=0.3]{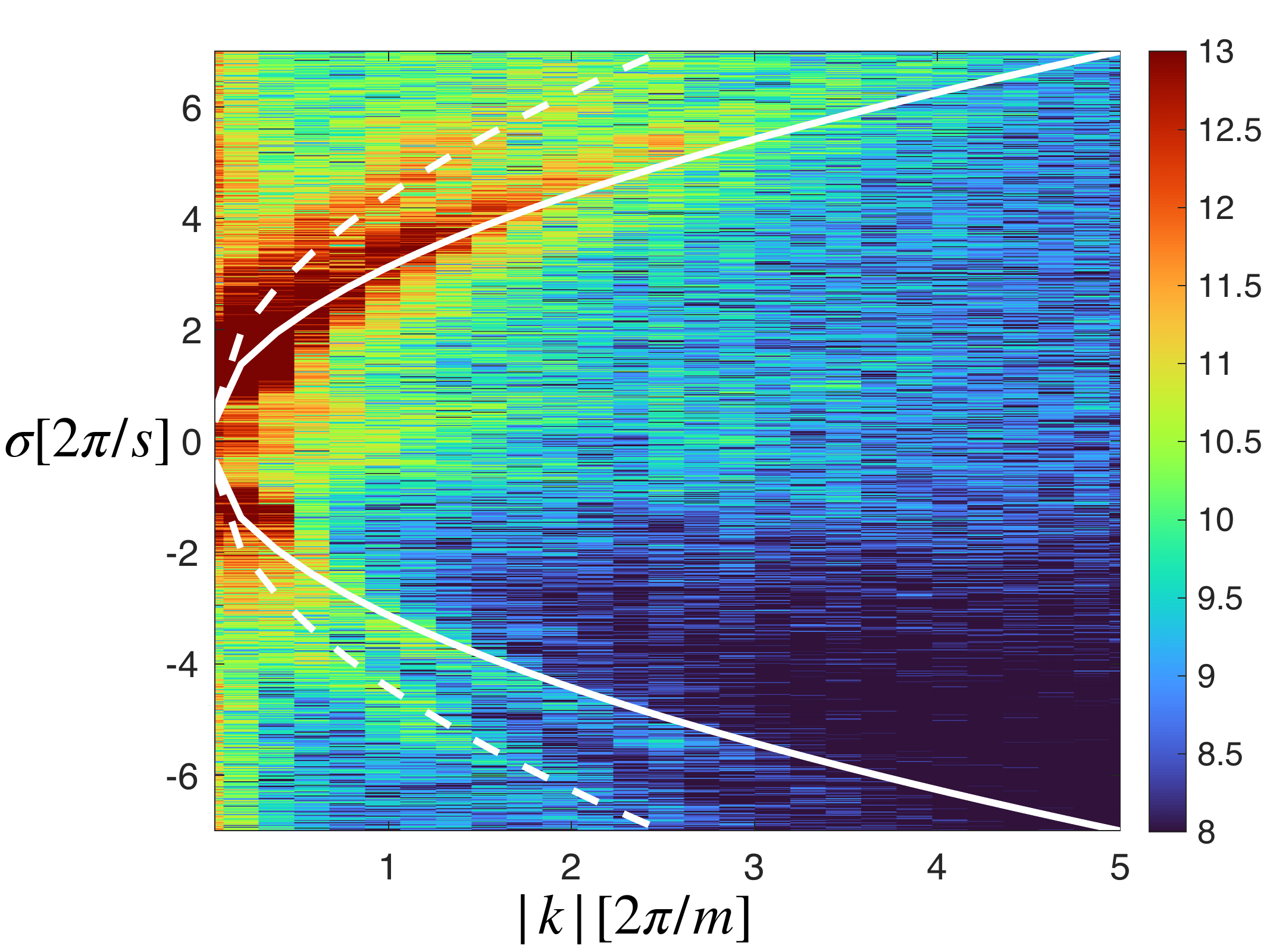} 
\caption{\label{fig:k-w_ww} 
Spatio-temporal spectrum of sea surface elevation measured at the Acqua Alta oceanographic tower, situated 15 km offshore from Venice. 
The red dashed lines correspond, starting from the top one, to  
$\sigma = \sqrt{2g|k|},  \sigma = \sqrt{g|k|}, \sigma =- \sqrt{g|k|},  \sigma =- \sqrt{2g|k|}$}.
\end{figure}
In this study, we derive the so-called normal variable $a(k,t)$ from the surface elevation field $\eta(x,y,t)$.
It is defined as 
\begin{equation}\label{eq:ak}
a(k,t)=\sqrt{\frac{g}{2\omega(k)}}\eta(k,t) - i \sqrt{\frac{\omega(k)}{2g}}\psi(k,t),
\end{equation}
where $k = (k_x, k_y)$ is a two-dimensional wavevector, $|k|$ denotes its magnitude and $\omega(k)=\sqrt{g |k|}$, with $g$ the gravity acceleration.   $\psi(k,t)$ is the velocity potential evaluated on the free surface, \cite{zakharov1999statistical,krasitskii1994}.
Since our measurements only provide the surface elevation and not the surface velocity, we estimate $\psi(k,t)$ using a leading-order approximation:
\begin{equation}
\label{eq:psi}
\psi(k,t) = \frac{g}{\omega(k)^2} \frac{d\eta(k,t)}{dt}.
\end{equation}
The rationale behind the introduction of the normal variable will be discussed in detail in Section~\ref{sec:waterwaves}.
Once the normal variable has been computed from the experimental data, we perform an additional Fourier transform in time to obtain $a(k,\sigma)$. 
The wave number-frequency spectrum is then given by $|a(k,\sigma)|^2$. 
To facilitate visualization, we convert from Cartesian coordinates $(k_x, k_y)$ to polar coordinates $(|k|, \theta)$ and integrate over the angular variable. The resulting spectrum is shown in Fig. \ref{fig:k-w_ww}.
The plot reveals the presence of multiple branches. The dominant one appears at positive frequencies and corresponds to free waves, closely following the linear dispersion relation $\sigma = \omega(k)=\sqrt{g|k|}$. As expected, a higher-order branch is also visible, which is well described by $\sigma = 2\omega(k/2)= \sqrt{2g|k|}$. This latter relation will be examined in detail in Section~\ref{sec:waterwaves}. Interestingly, a negative-frequency branch is also observed, with an intensity that is not symmetric with respect to its positive counterpart. An interpretation of these negative frequencies will be provided in the theoretical section of the paper.

\subsection{Numerical simulations of the water wave equations}\label{sec:numwater}
We supplement the experimental evidence in Section \ref{sec:expwater} with numerical simulations of water waves. In this section, we focus on the setup with a one-dimensional (1D) free surface, since the results are sufficient to reveal all key signatures on the spatio-temporal spectrum we study theoretically in this paper. The simulations with two-dimensional free surfaces have been conducted before, with the results of spatio-temporal spectra available in \cite{zhang2022numerical,zhang2025role}.

For gravity waves on a 1D free surface of an incompressible, inviscid and irrotational fluid, the flow can be described by a velocity potential $\phi(x,z,t)$ satisfying the Laplace's equation, with $x$ and $z$ the horizontal and vertical coordinates and $t$ the time. The surface velocity potential is defined as $\psi(x,t)=\phi(x,z,t)|_{z=\eta}$, where $\eta(x,t)$ is the surface elevation. The evolutions of $\eta$ and $\psi$ satisfy the Euler equations in Zakharov form \citep{zakharov1968stability}:
\begin{align}
    \eta_t+\eta_x\psi_x-(1+\eta_x^2)\phi_z=&0,
    \label{eq:eta}
    \\
    \psi_t + g\eta + \frac{1}{2}\phi_x^2-\frac{1}{2}(1+\eta_x^2)\phi_z^2&=0,
    \label{eq:phi}
\end{align}
where $\phi_z(x,t)=\partial\phi/\partial z|_{z=\eta}$ is the surface vertical velocity. 

We simulate \eqref{eq:eta} and \eqref{eq:phi} using the high-order spectral method \cite{dommermuth1987high} on a computational domain $[0,2\pi]$ with periodic boundary conditions and 512 grid points. We also re-scale mass and time so that both density $\rho$ and gravitational acceleration $g$ take values of unity, which we keep using for the rest of the section. The simulation includes up to cubic nonlinear terms in \eqref{eq:eta} and \eqref{eq:phi} through an order-consistent formulation \cite{west87,pan2020high}, allowing both triad and quartet interactions. The initial condition of the simulation is set as a realization of the JONSWAP spectrum \cite{hasselmann1973measurements} with peak wavenumber $k_p$=10, significant wave height $H_s=0.02$ (so that the effective steepness $k_pH_s/2=0.1$) and peak enhancement factor $\gamma$=6. More specifically, we take $\eta(k,t=0)$ as a realization of the spectrum with random phases and $\psi$ from the linear relation $\psi(k,t=0)=i\eta(k,t=0)/\omega(k)$. Such a configuration ensures that initially all waves are propagating to the negative $x$ direction. We conduct the simulation for 200$T_p$ with $T_p=2\pi/k_p^{1/2}$ the peak period, and compute the spatio-temporal spectrum $|a(k,\sigma)|^2$ in the time window $[180T_p,200T_p]$ using the same procedure described in Section \ref{sec:expwater} (except that $\psi(k,t)$ is now exactly available and the angle-averaging procedure is omitted).

The obtained spatio-temporal spectrum is plotted in figure \ref{fig:k-w_simwater}. In contrast to the measurement spectrum in figure \ref{fig:k-w_ww}, we preserve both positive and negative $k$'s in figure \ref{fig:k-w_simwater} due to the 1D nature of the simulation. We also see more branches in figure \ref{fig:k-w_simwater} than those in figure \ref{fig:k-w_ww} because of the higher precision in quantities and larger domain (relative to the peak wave length) enabled by the simulation. Five branches to be discussed are marked in figure \ref{fig:k-w_simwater} as the main branch and branches A-D. The main branch is the branch with highest intensity located on $\sigma=\omega(k)=\sqrt{g|k|}$ for negative $k$, which represents free waves traveling to the negative $k$ (or left) direction specified by the initial condition. Branch A, located on $\sigma=\omega(k)=\sqrt{g|k|}$ for positive $k$ with a weaker intensity, originates from free waves traveling to the positive $k$ (or right) direction. These waves are physically generated from quartet resonances with three modes of negative $k$ and one mode of positive $k$, as will be discussed in the water wave section below. Branch B corresponds to $\sigma=2\omega(k/2)=\sqrt{2g|k|}$, which is also identified in figure \ref{fig:k-w_ww} from measurement data. In addition, we observe a branch C located on the opposite side of the main branch compared to branch B, which is not shown in figure \ref{fig:k-w_ww} probably due to the limited observation domain at sea. Finally, we again find signals at negative frequencies (which is not possible for linear dynamics since the equation for canonical variable $a_k$ only permits positive $\sigma$) similar to that in figure \ref{fig:k-w_ww}. We mark the branch symmetric to and induced by the main brach as branch D (and not others for simplicity). A discussion on the formation of branches B-D, as well as some other negative-frequency branches, will be the topic of the theoretical part of this paper.  


\subsection{Numerical simulations of simplified models}
Although experimental data and full numerical simulations are extremely valuable, providing the ultimate test of a theory or revealing new physical phenomena, 
numerical simulations of {\it simplified models} remain a powerful tool: they offer high accuracy and allow the construction of minimal models that include only the  ingredients of interest. 
In this specific case, our focus is on investigating non-resonant interactions and identifying their signatures in the wavenumber-frequency spectrum.
We consider the following family of partial differential equations for the complex field $a(x,t)$:
\begin{equation}\label{eq:3waves}
    i \frac{\partial a}{\partial t} =
 \hat\omega(-i\nabla) \, a +\epsilon( a^2+2 aa^*),
\end{equation}
Here, $\hat\omega$ denotes a convolution (dispersive) operator. For our purposes, the operator must be chosen so that the equation does not permit any form of three-wave resonant interactions.
The parameter $\epsilon$ is small and controls the strength of the nonlinearity. The equation is solved using a pseudospectral method, where the dispersive term is treated in Fourier space, and time integration is performed with a fourth-order Runge-Kutta scheme. 
Periodic boundary conditions are imposed in a one-dimensional domain of length $L = 256$.
Although simulations were carried out for various dispersion curves, here we present only two representative cases corresponding to two distinct dispersion relations: $\omega(k)=\sqrt{g |k|}$ and $\omega(k)= |k|^2$.
\begin{figure}[h!]
\includegraphics[scale=0.48]{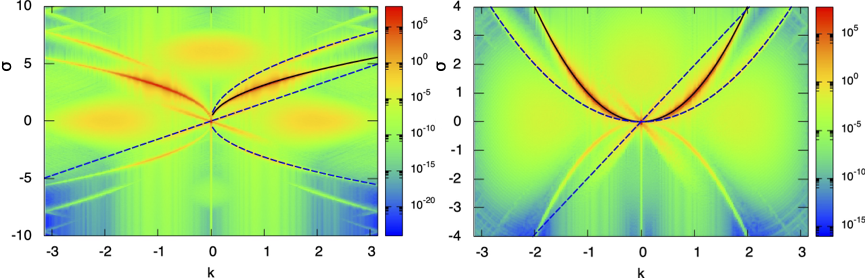} 
\caption{\label{fig:num_k_w} The spatio-temporal spectra obtained from numerical simulations of the equation in 
\eqref{eq:3waves} with dispersion relation $\omega(k)=\sqrt{g|k|}$ (left plot) and $\omega(k)=|k|^2$ (right plot). 
Both plots show different excitations: the solid black curves in each plot corresponds to the dispersion relation for the free waves, while blue dashed lines are attributed to bound modes. In the left plot, the equation for the bound modes are $\sigma(k)=\pm \sqrt{2 g |k|}$ and $\sigma(k)=c_g k$, where $c_g$, is the group velocity computed at the peak of the spectrum, $k\simeq 1$.  Similarly, in the right plot, the equation for the bound modes are $\sigma(k)=|k|^2/2$ and $\sigma(k)=c_g k$, where $c_g$, is the group velocity computed at the peak of the spectrum, $k\simeq 1$.}
\captionsetup{justification=raggedleft,singlelinecheck=false}
\end{figure}
The first mimics surface gravity waves (with $g$ denoting gravitational acceleration), while the second corresponds to a Schr\"odinger-type dispersion.
Although the equations obtained may exhibit pathological behavior at long times, we restrict our numerical simulations to a time window that allows us to analyze the excitations in the spatio-temporal spectrum. 
Initial conditions are provided in Fourier space with a spectrum $|a_k(t=0)|^2$ given by the superposition of two Gaussian shape functions centered at $k\simeq 1$ and $k\simeq-1$. The Fourier phases are taken uniformly distributed in the interval $(0,2\pi]$. 
A space-time Fourier transform of the solution is then performed, and the quantity $|a(k,\sigma)|^2$ 
is computed and plotted in Fig. \ref{fig:num_k_w}. Both cases show that the spectrum is mostly energetic on the lines corresponding to the linear dispersion relation. Higher order bound modes, including negative frequencies,  are also evident. In what follows, we provide a theoretical explanation of all the branches depicted in the figure.




\medskip
\section{Space resonances in simplified models} \label{sec:space_res}

\vspace{0.5em}
We begin by considering a homogeneous dispersive nonlinear PDE for a complex field $a(x,t)$ defined on the $d$-dimensional torus $\mathbb{T}_L^d$ with periodic boundary conditions. In Fourier space, the dynamics is governed by a Hamiltonian of the form: 
\begin{align} \label{eq:Hsimple}
\mathcal{H} &= \sum_{k} \omega(k) a_k a_k^* + \epsilon \sum_{k_1,k_2,k_3} V_{k_2 k_3}^{k_1} (a_1^* a_2 a_3 + a_1 a_2^* a_3^*) \delta_{k_2 k_3}^{k_1},
\end{align}
where $k \in \mathbb{Z}_L^d := \mathbb{Z}^d / L$, and the Fourier amplitudes $a_k$ are complex-valued. The dispersion relation $\omega(k)$ is real, $V^{k_1}_{k_2,k_3}$ denotes the interaction coefficient 
and $\delta_{k_2 k_3}^{k_1}$ is the Kronecker delta that is equal to 1 if $k_1=k_2+k_3$  and 0 otherwise. 
The small parameter $0 < \epsilon \ll 1$ controls the strength of nonlinearity. 
We omit factors of $L^{-d}$ in the sums for light notation,
\and will sometimes denote $\omega_k=\omega(k)$ and $\omega_j = \omega(k_j)$ when $j=1,2,3,\dots$,
omit commas in the indexes of the coefficients, e.g. $V^{k_1}_{k_2,k_3} = V^{k_1}_{k_2k_3}$,
and write $V^{k}_{k_2,k_3} = V^k_{23}$ or $V^{k_1}_{k_2,k_3} = V^1_{23}$ and so on. Note that the case $d=1$ and $V^{k_1}_{k_2k_3}=1$ corresponds to \eqref{eq:3waves}.

From \eqref{eq:Hsimple}, the equations of motion are
\begin{align}\label{eq:Hsimplemotion}
\partial_t a_k &= -i \frac{\partial \mathcal{H}}{\partial a_k^*} = -i \omega_k a_k 
- i\epsilon \sum_{k_2,k_3} V_{k_2,k_3}^k a_2 a_3 \delta_{k_2 k_3}^k 
- 2i\epsilon \sum_{k_2,k_3} V^{k, k_3}_{k_2} a_2 a_3^* \delta_{k k_3}^{k_2} 
\end{align}
where we are setting $V^{k, k_3}_{k_2} = V_{k, k_3}^{k_2}$ adopting the convention
that the output frequency $k$ is always an upper indexes 
and that the other upper indexes in the coefficients correspond 
to variables that have a complex conjugate, like $a_3^\ast$.

In the linear case $\epsilon = 0$, the system supports free dispersive waves $a_k(t) = a_k^0 e^{-i\omega_k t}$. To study nonlinear interactions, we define the slowly varying envelope 
\begin{align}\label{eq:b_env}
    b_k(t) := a_k(t) e^{i\omega_k t},
\end{align}
which filters the fast linear oscillation: 
\begin{align}\label{dtbk}
\partial_t b_k &= -i\epsilon \sum_{k_2,k_3} V_{k_2k_3}^k b_2 b_3 
e^{i(\omega_k-\omega_2 -\omega_3)t} \delta_{k_2 k_3}^k 
  - 2i\epsilon \sum_{k_2,k_3} V^{k k_3}_{k_2} b_2 b_3^* 
  e^{i(\omega_k-\omega_2 +\omega_3)t} \delta_{k k_3}^{k_2} + \mathcal{O}(\epsilon^2, b^3).
\end{align}
Integrating in time yields the Duhamel representation:
\begin{align} \label{eq:b}
b_k(t) = b_k^0 &- i\epsilon \sum_{k_2,k_3} V_{k_2k_3}^k \int_0^t b_2(s) b_3(s) 
  e^{i(\omega_k-\omega_2 - \omega_3)s} ds \, \delta_{k_2 k_3}^k 
\\
&- 2i\epsilon \sum_{k_2,k_3} V^{k k_3}_{k_2} \int_0^t b_2(s) b_3^*(s) 
  e^{i( \omega_k-\omega_2 + \omega_3)s} ds \, \delta_{k k_3}^{k_2} + \mathcal{O}(\epsilon^2, b^3), \nonumber
\end{align}
which makes explicit how nonlinear interactions among various modes accumulate over time and contribute to the evolution of $b_k(t)$. Two key concepts arise:

\vspace{0.5em}
\noindent\textbf{Concept 1: Time Resonances}.
Whenever the interaction phase $$\omega^k_{23} := \omega_k - \omega_2 - \omega_3 \neq 0,$$
we can then integrate by parts in time. Using the notation
\begin{align*}
\Delta(t;x) = 
\left\{ \begin{array}{cc}
     \dfrac{e^{ix t} - 1}{ix} & x\neq 0,  
     \\
     \\
     t &  x = 0,
\end{array}
\right.
\qquad \Big( \frac{d}{dt} \Delta(t;x) = e^{itx} \Big)
\end{align*}
and the fact that $\partial_s (b_2 b_3) = \mathcal{O}(\epsilon)$, we obtain
\begin{align} \label{eq:first_nl}
\int_0^t b_2(s) b_3(s) e^{i\omega^k_{23}s} ds = 
\Delta(t;\omega^k_{23})  
\, b_2^0 b_3^0 + \mathcal{O}(\epsilon);
\end{align}
plugging \eqref{eq:first_nl} (and its analogue with the phase 
$\omega^{k3}_2 := \omega_k - \omega_2 + \omega_3$ assuming that that also does not vanish) into \eqref{eq:b}, we derive
the leading-order solution:
\begin{align}\label{eq:b_leading}
\begin{split}
b_k(t) & = B_k^{(0,1)}(t) + \epsilon B_k^{(1,1)}(t) + \mathcal{O}(\epsilon^2), 
\\
B_k^{(1,1)}(t) & := - \sum_{k_2,k_3} B^k_{k_2k_3}
  e^{i\omega^k_{23} t} \delta_{k_2 k_3}^k - 2 \sum_{k_2,k_3} B^{kk_3}_{k_2} 
  e^{i\omega^{k3}_2 t} \delta_{k k_3}^{k_2},
\end{split}  
\end{align}
where we are introducing the notation
\begin{equation}\label{eq:b_leadingcoeff}
B^k_{k_2k_3} := \frac{V_{k_2k_3}^k}{\omega^k_{23}} b_2^0 b_3^0,
\qquad B^{kk_3}_{k_2} := \frac{V^{kk_3}_{k_2}}{\omega^{k3}_2} b_2^0 (b_3^0)^\ast,
\end{equation}
and let
\begin{equation}\label{eq:b_leadingcoeff0}
B_k^{(0,1)} := b_k^0 + \epsilon \sum_{k_2,k_3} B^k_{k_2k_3}  \delta_{k_2 k_3}^k
  + 2\epsilon \sum_{k_2,k_3} B^{kk_3}_{k_2} \delta_{k k_3}^{k_2}.
\end{equation}

The integration by parts procedure leading to \eqref{eq:b_leading}, 
which relies on the non-vanishing of the interaction phase, 
is equivalent to a classical (Poincar\'e) normal form transformation 
for the system of ODEs \eqref{eq:Hsimplemotion}. 
In essence, this procedure removes the quadratic nonlinear terms from the dynamics, 
yielding a more accurate approximation of the solution in the perturbative regime. 
For PDEs (viewed as infinite-dimensional systems of ODEs), 
the concept of normal forms was introduced and formalized by Shatah \cite{shatahKGE}. 
Since then, it has become an essential tool in the rigorous analysis of PDEs, 
particularly in applications to water wave theory; 
see, for instance, \cite{GMS2}, \cite{ADa}, \cite{IoPu4}, \cite{BFP}, and \cite{DIP1}.

The above calculation motivates the definition of the classical concept of resonance:

\vspace{0.5em}

\noindent\textbf{Definition 1 (Time resonance).} A triad $(k, k_2, k_3)$ satisfies a time resonance if
\begin{equation}\label{timeres} 
k = k_2 + k_3 \quad \text{and} \quad \omega(k) = \omega(k_2) + \omega(k_3).
\end{equation}
Similarly if $k = k_2 - k_3$ and $\omega(k) = \omega(k_2) - \omega(k_3)$.

In the absence of time resonances (e.g., when $\omega(k)$ is concave), the perturbative expansion \eqref{eq:b_estimate} remains valid over long times. 


\subsection*{The spatio-temporal spectrum}
From \eqref{eq:b_leading}, it is straightforward to derive the leading order contribution to the spatio-temporal spectrum of $a_k$: 
transforming back $b_k=a_ke^{i\omega_kt}$ and taking the Fourier transform in time 
(note our unconventional choice of sign for the exponential which is done for convenience)
$$\tilde{a}_{k}\left(\sigma\right):=\int_{-\infty}^{\infty}a_{k}\left(t\right)e^{i\sigma t}dt,$$ 
we arrive at 
\begin{align}\label{eq:spatio_amplitude}
\begin{split}
\tilde{a}_{k}\left(\sigma\right) = B_{k}^{\left(0,1\right)}
  \delta\left(\sigma-\omega_{k}\right)
  & - \epsilon \sum_{k_{2},k_{3}} B^{k}_{k_2k_3}
\delta_{k_{2}k_{3}}^{k}\delta\left(\sigma-\omega_{1}-\omega_{2}\right)
\\
  & - 2\epsilon\sum_{k_{2},k_{3}} B^{kk_3}_{k_2}
  \delta_{kk_{3}}^{k_{2}}
  \delta\left(\sigma-\omega_{2}+\omega_{3}\right) + O\left(\epsilon^{2}\right),
\end{split}
\end{align}
having used that $b_k^0=a_k^0$.
If we assume that the initial data is Gaussian with the second moment given by
\begin{align}\label{nk}
    \left\langle a_{k}^{0}a_{p}^{0}\right\rangle &=n_{k}\delta_{kp},
\end{align}
where $\left\langle \, \cdot \, \right\rangle$ denotes average with respect to the initial data distribution, 
we can write a closed expression for the spatio-temporal spectrum at order $\epsilon^2$.  
Assuming the system is in a (time and space homogeneous)
statistically steady state 
\begin{equation}\label{Sk}
    \left\langle \tilde{a}_{k}\left(\sigma\right)\tilde{a}_{p}^{*}\left(\sigma_{1}\right)\right\rangle =S_{k}(\sigma)\delta_{kp}\delta\left(\sigma-\sigma_{1}\right),
\end{equation}
using \eqref{eq:spatio_amplitude} we arrive at
\begin{align}\label{eq:spatiotemp_simple}
\begin{split}
S_{k}(\sigma) = \mathcal{N}_{k}\delta\left(\sigma-\omega_{k} + \epsilon^{2}\delta\omega_{k}\right)
& +\epsilon^{2}\sum_{k_{2},k_{3}}\left\langle \left|B^{k}_{k_2k_3}
  \right|^{2}\right\rangle \delta_{kk_{3}}^{k_{2}}\delta\left(\sigma-\omega_{2}-\omega_{3}\right)
 \\
 & + 4\epsilon^{2}\sum_{k_{2},k_{3}}\left\langle \left| B^{kk_3}_{k_2} 
 \right|^{2}\right\rangle \delta_{kk_{3}}^{k_{2}}\delta\left(\sigma-\omega_{2}+\omega_{3}\right)+O\left(\epsilon^{4},a^{4}\right),
\end{split}
\end{align}
where:

\setlength{\leftmargini}{1.5em}
\begin{itemize}
\item[-] The Gaussian average gives 
\begin{align}\label{coeffave}
\left\langle \left| B^{k}_{k_2k_3}
  \right|^{2}\right\rangle =\left(\frac{V_{23}^{k}}{\omega_{23}^{k}}\right)^{2}n_{2}n_{3},
  \qquad 
  \left\langle \left| B^{kk_3}_{k_2}
  \right|^{2}\right\rangle =\left(\frac{V^{k3}_{2}}{\omega_{2}^{k3}}\right)^{2}n_{2}n_{3};
\end{align}

\item[-] The frequency shift $\delta\omega_k$ is 
due to four-wave resonant terms arise from the second order expansion $a_k^{(2)}$, that are included in the $\mathcal{O}(\epsilon^2)$ in \eqref{eq:b_leading};
these are discussed more in details later in this section; 

\item[-] The correction to the second order moment is given by 
\begin{align}\label{Nk}
\begin{split}
\mathcal{N}_{k} & := \left\langle \left| B_{k}^{(0,1)} \right|^{2}\right\rangle
  = n_{k}+\epsilon^{2}\sum_{k_{2},k_{3}}\left(\frac{V_{23}^{k}}{\omega_{23}^{k}}\right)^{2}n_{2}n_{3}
  \delta_{kk_{3}}^{k_{2}}
  + 4\epsilon^{2}\sum_{k_{2},k_{3}}\left(\frac{V^{k3}_{2}}{\omega^{k3}_{2}}\right)^{2} n_{2}n_{3}
  \delta_{kk_{3}}^{k_{2}}.
\end{split}
\end{align}


\end{itemize}

Note that, at $\epsilon=0$, the equations are linear and the spatio-temporal spectrum is concentrated on the linear branch $\delta(\sigma-\omega_k)$. As $\epsilon\neq 0$, {\it non-resonant nonlinear} 
interactions contribute additional branches to the spatio-temporal spectrum given by the $\epsilon^2$ terms in \eqref{eq:spatiotemp_simple} on
 which we now concentrate.

\begin{figure}[h!]
\includegraphics[scale=0.23]{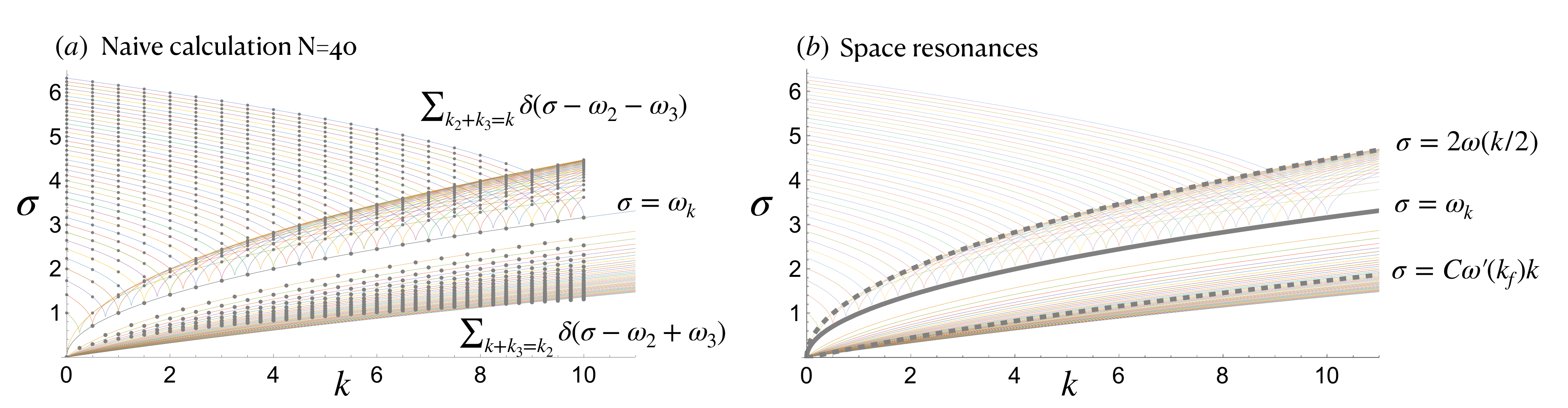} 
\caption{\label{fig:spatiosimpletheory} 
(a) Naive approximation of the theoretical spatio-temporal spectrum $S_k$ 
from the simplified model, showing all triadic branches, assuming the coefficients $A^{(i,j)}=1$, $\omega_k = \sqrt{g |k|}$ with $g=1$ and resolution $L = 40$. In other words, we plot the support set of the spatio-temporal spectrum. Specifically, above the $\sigma=\omega(k)$ curve, $\sum_{k_2+k_3=k} \delta(\sigma -\omega(k_2)-\omega(k_3))$ is plotted: for each $k$ we plot $\sum _{k_2}\mathcal{\chi}_{\sigma=\omega(k-k_2)-\omega(k_2)}$ where $\mathcal{\chi}$ is the indicator function and $k_2$ runs over the wavenumbers on the lattice. The dots corresponds to all values of $\sigma$ where the indicator function does not vanish. The curves of different colors correspond to the sets of support of the indicator functions in the sum in the continuum limit $L\rightarrow\infty$. This is equivalent to a naive approximation of the normal form transformation. 
(b) The space resonances of the leading order expansion and the linear dispersion relation.}
\end{figure}

We begin by noting that the expressions \eqref{eq:spatio_amplitude} and \eqref{eq:spatiotemp_simple} must be interpreted with care. At first glance, they suggest that all nonresonant interactions at order $\epsilon$ contribute more or less equally to the spatio-temporal spectrum of the variable $a$, leading to a broad distribution supported on all additional branches generated by three-wave interactions, for example $\sum_{k_2+k_3=k} \delta(\sigma - \omega_2-\omega_3)$.
However, numerical simulations and observational data show that only a subset of these interactions dominate. 
 To illustrate, consider the contribution of the nonlinear terms in the spatio-temporal spectrum given by \eqref{eq:spatiotemp_simple} with domain size $L=1/4$ and $\omega_k=\sqrt{g|k|}$, and assume the coefficients in \eqref{coeffave} are $O(1)$. The resulting spectrum is shown in Figure \ref{fig:spatiosimpletheory} (a).  
 
\vspace{0.5em}
\noindent\textbf{Concept 2: Space Resonances.}
A closer look at \eqref{eq:spatiotemp_simple} shows that the contribution of non-resonant interactions to the spatio-temporal, given by
\[
\sum_{k_{2},k_{3}}\left\langle \left|B^{k}_{k_2k_3}\right|^{2}\right\rangle 
\delta_{kk_{3}}^{k_{2}}\delta\left(\sigma-\omega_{2}-\omega_{3}\right),
\]
tend to cluster at specific locations in the spectrum;
see Figure \eqref{fig:spatiosimpletheory}(b).
While each individual contribution is small, their accumulation leads to enhanced energy concentration at those locations. The largest contributions occur when the argument of $\delta\left(\sigma-\omega_{2}-\omega_{3}\right)$ becomes stationary with respect to changes in the wave number. This observation motivates the following definition:

\begin{definition}[Space Resonance]
Under the constraint $k = k_2 + k_3$, a three-wave interaction is space-resonant if the phase $\omega_{23}^k$ is stationary:
\begin{align}\label{SR}
\nabla_{k_{2}}(\omega^{k}_{23}) \, \delta^k_{23}
  = - \nabla\omega\left(k_{2}\right) + \nabla\omega\left(k-k_{2}\right) & = 0.
\end{align}
\end{definition}

\smallskip
As mentioned in the introduction, the absence of space resonances plays a key role in proving long-time existence and scattering  results in nonlinear dispersive PDEs, see for example \cite{germain2009global,GMSC};
the simultaneous presence of space and time resonances instead, often leads to nonlinear leading order dynamics
\cite{KP,DIPP,DPWave} or may even cause blow-up. 
In this work, differently from the case of small data scattering, we give evidence that space resonances influence the spatio-temporal spectrum, and give rise to nonlinear features also in the absence of time resonances.

\smallskip
Another way that explains why some nonresonant triads are selectively enhanced, we analyze the long-time behavior of \eqref{eq:b_leading}. In the continuum limit ($L \to \infty$), the sums over discrete modes converge to integrals:
\begin{equation}
\label{sumtoint}
\frac{1}{(2\pi L)^d}
\sum_{k_2, k_3 \in (\Z/L)^d} F(k,k_2,k_3) \, \delta^k_{k_2 k_3} 
    \to \int F(k,k_2,k_3) \, \delta(k - k_2 - k_3) \, dk_2 dk_3 \,.   
\end{equation}
If the interaction phase is not stationary, i.e.\ $\nabla_{k_2}(\omega^k_{23} ) \, \delta^k_{23} \neq 0$, one can integrate by parts to show that such contributions decay at long times. Specifically,
\[
i t e^{i \omega^k_{23} t}= \frac{\nabla_{k_2} \omega^k_{23}}{|\nabla_{k_2} \omega^k_{23}|^2} 
  \cdot \nabla_{k_2} e^{i \omega^k_{23} t},
\]
and integrating by parts in $k_2$ gives
\begin{align}
-\epsilon \int dk_2 dk_3\, B^{k}_{k_2 k_3}
  e^{i \omega^k_{23} t} \delta^k_{k_2 k_3}
= \frac{\epsilon}{it} \int dk_2 dk_3\, \nabla_{k_2} \cdot \left( B^{k}_{k_2 k_3}
\frac{\nabla_{k_2} \omega^k_{23}}{|\nabla_{k_2} \omega^k_{23}|^2} \right) 
  e^{i \omega^k_{23} t} \delta^k_{k_2 k_3},
\end{align}
where we neglect boundary terms for simplicity. The additional factor of $1/t$ highlights the decay of non-stationary contributions and motivates the space resonance condition \cite{germain2009global}.

Condition \eqref{SR} is equivalent to requiring that the group velocities of the interacting waves coincide, i.e., $\nabla \omega(k_2) = \nabla \omega(k - k_2)$. Physically, this means that the corresponding wave packets remain co-located in space, allowing coherent energy exchange over long times even without exact frequency resonance. Geometrically, the condition \eqref{SR}, or more generally the smallness of the gradient, corresponds to the regions where the curves in Figure \ref{fig:spatiosimpletheory} become denser.

Combining concepts 
1 and 2, we conclude that interactions which are neither time nor space resonant
are more perturbative and decay 
faster than linear solutions for 
long times.


\subsection*{Space Resonances signatures in the spectrum}
In systems with concave dispersion relations, such as surface gravity waves with $\omega(k) = \sqrt{g|k|}$,
exact three-wave time resonances are absent, 
aside from the trivial cases that involve a zero-mode (in the input or output);
to avoid complications from the possible presence of small divisors introduced by these trivial resonance,
we assume, for the moment, that the interaction coefficients vanish when one of the wave numbers is zero;
a more precise discussion of zero frequencies will be given later on.

Three-wave space resonances instead, always exist, for instance at $k_2 = k_3 = k/2$. This corresponds to a self-interaction of the wave number $k/2$ with itself. Retaining only the leading-order contributions from space-resonant triads, we obtain the crude approximation for large times ($T \omega_k \gg 1$):
\begin{align} \label{eq:b_estimate}
b_k(T) \sim B_k^{(0,1)} 
- \epsilon B^{k}_{k/2, k/2} \, e^{i \omega^k_{k/2,k/2} T}
- \lim_{k \to 0} 2\epsilon \sum_{k_2} B^{k,k_2-k}_{k_2} 
    \, e^{i \omega^{k,k_2-k}_{k_2} T}
+ O\left(\frac{\epsilon}{T}\right). 
\end{align}
The first nonlinear term corresponds to the space-resonant triad $(k, k/2, k/2)$, 
which yields a peak in the spatio-temporal spectrum at
\begin{equation}\label{harmonic}
\sigma = 2\omega(k/2). 
\end{equation}
The second nonlinear term in \eqref{eq:b_estimate} captures interactions with 
slowly varying modes and includes space-resonant triads of the form $(k, k_2-k, k_2)$ when $k \approx 0$. 
Indeed, the space resonance condition for the second term is
\begin{align}\label{eq:2space}
\nabla_{k_2} \left( \omega(k) - \omega(k_2) + \omega(k_2-k) \right) 
= -\nabla\omega(k_2) +  \nabla \omega(k_2-k) = 0,
\end{align}
which is satisfied for all $k_2$ when $k = 0$. 
If the energy spectrum $n_k$ is peaked near a characteristic wave number $k_f$, the dominant contribution arises from $k_2 = k_f$. 
Expanding the phase for small wave-number $k$ as 
\begin{align}\label{cgk}
\omega(k_f-k) - \omega(k_f) \approx -\omega'(k_f) k,
\end{align}
reveals a 
peak in the spatio-temporal spectrum at the curve 
\begin{align}\label{Doppler}
\sigma = \omega'(k_f) k. 
\end{align}
For larger $k$ the approximation on the right-hand side of \eqref{cgk} is not accurate, and more terms need to be included; 
see the fit for branch C in Figure \ref{fig:k-w_simwater}.
It is important to note that both concentration curves \eqref{harmonic} and \eqref{Doppler} 
may be suppressed (or even enhanced) by the values of the symbols at the respective interacting triads.

In summary, already for the simple model \eqref{eq:Hsimplemotion}, two types of space-resonant peaks are expected: 
$\sigma = 2\omega(k/2)$, from self-interactions of mode $k/2$,
and $\sigma = \sign(k_f) \omega'(|k_f|) k$, from interactions with slowly varying background or large-scale flow.
The signatures of these interactions are evident in the numerical and experimental data, as shown in 
Figures  \ref{fig:k-w_ww}, \ref{fig:k-w_simwater} and \ref{fig:num_k_w}.

\begin{remark}
For a slightly more accurate description compared to \eqref{eq:b_estimate},
one can use the method of stationary phase which, in the continuum limit,  gives
\begin{align}
& \epsilon \sum_{k_2, k_3} B^{k}_{k_2 k_3} e^{i \omega^k_{23} t} \delta^k_{k_2 k_3} 
\\
& \sim \epsilon C e^{i \omega(k) t} B^{k}_{k/2, k/2} e^{-i 2\omega(k/2) t} \left( \frac{2\pi}{t} \right)^{d/2}
  \frac{1}{\sqrt{|\det D^2 \phi(k/2)|}} + O(t^{-(d+1)/2}),
\end{align}
where $C$ is a constant. 
\end{remark}


\medskip
\vspace{0.5em}
\noindent
\textbf{Heuristic Space Resonances: the Continuum Limit.}
As an additional confirmation of the peaks observed above, 
we evaluate formally the spatio-temporal spectrum in the continuum limit. 
Consider one of the leading non-resonant terms in \eqref{eq:spatiotemp_simple}
\begin{align}\label{Acontinuous}
A 
  (\sigma, k) := \epsilon^2 \int dk_2 dk_3\, \left\langle |B^{k}_{k_2k_3} 
  |^2 \right\rangle
  \delta(k - k_2 - k_3) \delta(\sigma - \omega_2 - \omega_3).
\end{align}
Integrating over $k_3$ together with $\delta(k-k_2-k_3)$ gives $k_3=k-k_2$. 
Then, recall that 
\begin{align}
    \int g(k_2)\delta(f(k_2))dk_2 = 
    \int_{\{f=0\}} g \, dS= \sum_{k_2^\star} \int g(f^{-1}(k_2)) \frac{\delta(k_2-k_2^\star)}{|\nabla f(k_2)|}dk_2,
\end{align}
where $k_2^\star$ are the roots $f(k_2^\star)=0$. 
Integrating over $k_2$ with the remaining delta function $\delta(f(k_2))$ 
with $f(k_2)\equiv\sigma-\omega(k_2)-\omega(k-k_2)$
reduces the integral to a sum over the roots $k_2^\star$ 
satisfying $\omega(k_2^\star) + \omega(k - k_2^\star) = \sigma$:
\begin{align}\label{eq:exact_spatio}
A 
(\sigma, k) 
= \sum_{k_2^\star} \left\langle |B^{k}_{k_2^\star, k - k_2^\star}|^2 \right\rangle 
\frac{1}{|\nabla_{k_2} \left( \omega(k_2) + \omega(k - k_2) \right) |} 
 \, \Big|_{k_2 = k_2^\star}.
\end{align}
Sharp peaks in the spectrum arise when the derivative in the denominator becomes small; and the vanishing of the denominator is precisely the condition for space resonance which results in the $\delta$ contributions derived above.

For 1D surface gravity waves, $\omega(k) = \sqrt{g|k|}$, the resonant roots and derivatives are explicit:
\[
k_{2,\pm}^\star = \frac{g|k| \pm \sqrt{2g|k| \sigma^2 - \sigma^4}}{2g},
  \qquad \big( \, k_{2,\pm}^\star = k/2 \quad \Longleftrightarrow \quad \sigma = \sqrt{2g|k|} \, \big).
\]



\subsection*{Nonlinear Frequency Shift from Four-Wave Interactions} 
Before presenting a leading-order approximation of the spatio-temporal spectrum
we first examine the contribution of the $\mathcal{O}(\epsilon^2)$ terms in the expansion of the amplitude $b(t)$ from \eqref{eq:b_leading} 
. The next order expansion, $\mathcal{O}(\epsilon^2)$, of the amplitude contributes to the spatio-temporal spectrum at this order only to the shift of the linear phase. Specifically these are terms of the form $(b,b,b^\ast)$. 
Performing one additional integration by parts on the residual terms from \eqref{eq:b_leading}, we obtain a representative (gauge-invariant) four-wave interaction term as in the following:
\begin{align}\label{phaseshift0}
b(t) = B_k^{0,1} + \epsilon B_k^{(1,1)} + 4\epsilon^2 \sum_{k_2, k_3} \sum_{k_4, k_5} 
  \left( V_{k_2k_5}^{k_4} \frac{V_{k_2k_3}^{k}}{ \omega_{23}^{k}} + \cdots \right) b_3^0 b_4^0 b_5^{0*} 
  \frac{e^{i \omega_{43}^{k5} t} - 1}{\omega_{43}^{k5}}
  \delta_{k_2 k_3}^{k} \delta_{k_2 k_5}^{k_4} \\ + \mathcal{O}_\text{Wick}(\epsilon^2) + \mathcal{O}(\epsilon^3),
\end{align}
where $\omega_{43}^{k5} := \omega_k + \omega_5 - \omega_4 - \omega_3$;
the terms $B_k^{(0,1)}$ and $B^{(1,1)}_k$ are defined in
\eqref{eq:b_leading} and \eqref{eq:b_leadingcoeff0}
and we will not keep track of them in this subsection as they 
do not contribute to the phase shift. 
while $\mathcal{O}_\text{Wick}(\epsilon^2)$ denotes terms that vanish under Wick contraction 
in the computation of the spatio-temporal spectrum at order 
$\mathcal{O}(\epsilon^2)$.
Also, note that 
we are not writing out explicitly all the symbols for the sake of discussion, but
we will detail these terms for the water waves system at the end of Section \ref{sec:waterwaves}.

Among the nonlinear contributions in \eqref{phaseshift0}, most 
terms 
do not contribute to the spectrum at order $\mathcal{O}(\epsilon^2)$,
as they vanish under Wick contraction
unless they correspond to {\it trivial} 
four-wave time resonances, 
i.e., 
$k_3=k_5$ or $k_4=k_5$.
In that case, using the limit identity \( \lim_{x \to 0} \frac{e^{itx} - 1}{x} = it \), 
we obtain a leading-order nonlinear phase shift:
\begin{align}\label{bshift}
b(t) = B_k^{(0,1)} + \epsilon B^{(1,1)}(t) + it  4\epsilon^2 b_k^0  \, \sum_{k_3} 
   \left( \delta_{k_2 k_3}^{k} \sum_{k_2} \frac{(V_{k_2k_3}^{k})^2}{\omega_{23}^{k}} + \cdots \right) b_3^0 b_3^{0*} \,  
  + \cdots.
\end{align}
Note that the third term on the right-hand side of \eqref{bshift} has the form $$i t 4\epsilon^2 b_k^0  \, \sum_{k_3} M_{k,k_3} \, b_3^0 b_3^{0*}$$ for some real valued symbol $M_{k,k_3}$.  
This apparent secular growth actually contributes to a nonlinear frequency correction to the mode \( b_k \), 
and perturbs the support of the linear frequency branch in the spatio-temporal spectral amplitude.
Indeed, taking the Fourier transform in time, the phase modulation term proportional to \( it \) contributes a derivative of the Dirac delta:
\begin{align}
\int_{-\infty}^{\infty} dt\, it\, e^{-i\omega_k t + i \sigma t} 
    =: \delta'(\sigma - \omega_k),
\end{align}
so that the Fourier-transformed amplitude becomes (recall \eqref{eq:b_leadingcoeff0})
\begin{align}\label{eq:a_freq}
\tilde{a}_k(\sigma) = 
B_k^{(0,1)} \delta(\sigma - \omega_k) + \mathcal{O}(\epsilon) 
    + \epsilon^2 \sum_{k_3} 
  \Big( \sum_{k_2} A_{k k_2 k_3}^{(1,2)}\delta_{k_2 k_3}^{k} + \cdots \Big) a_k^0 \delta'\left(\sigma - \omega_k\right) + \cdots,
\end{align}
where (cfr. with \eqref{eq:b_leadingcoeff} and \eqref{bshift})
$
A_{k k_2 k_3}^{(1,2)} := \frac{(V_{k_2k_3}^{k})^2}{\omega_{23}^{k}} b_3^0 b_3^{0*} = 
    V_{k_2k_3}^{k} B_{k_2k_3}^{k}
$
and ``$\cdots$'' denote other symbols that we do not write explicitly here.
Combining terms, and comparing with a Taylor expansion of the $\delta$ 
(in the 
sense of duality with the space of smooth functions of compact support), 
we can express the result as a ``shifted delta'': 
\begin{align}
\tilde{a}_k(\sigma) = B_k^{(0,1)} \delta\left(\sigma - \omega_k + \frac{\epsilon^2}{B_k^{(0,1)}} 
  \sum_{k_3} \left( \sum_{k_2} A_{k k_2 k_3}^{(1,2)} \delta_{k_2 k_3}^{k} + \cdots \right) a_k^{0} \right) 
  + \mathcal{O}(\epsilon) + \cdots.
\end{align}
This frequency shift appears directly in the spatio-temporal spectrum,
as anticipated in \eqref{eq:spatiotemp_simple}:
\begin{align}
S_k(\sigma) = \mathcal{N}_k \delta\left(\sigma - \omega_k + \delta\omega_k \right) + \mathcal{O}(\epsilon) + \cdots,
\end{align}
where the nonlinear frequency correction from this representative term is given by
\begin{align}\label{eq:phase_shift}
\delta\omega_k = \epsilon^2 
\sum_{k_2, k_3} \left\langle A_{k k_2 k_3}^{(1,2)} \right\rangle \delta_{k_2 k_3}^{k} + \cdots.
\end{align}

\subsection*{Leading-Order Approximation of the Spectrum}
We can now summarize the dominant contributions to the spectrum from space resonances:
\begin{align}\label{eq:finite_spectrum}
S_k(\sigma) &\approx \mathcal{N}_k \delta\left( \sigma - \omega_k + \epsilon^2 \delta \omega_k \right) \\
&\quad + \epsilon^2 \left\langle | B^{k}_{k/2, k/2} |^2 \right\rangle \delta\left( \sigma - 2\omega(k/2) \right) 
  + 4 \epsilon^2 
  \left\langle |B^{k, k_f - k}_{k_f} |^2 \right\rangle \delta\left( \sigma - \omega'(k_f) k \right) 
  + O(\epsilon^4). \nonumber
\end{align}
This form  is peaked along the linear branch, a ``second-harmonic'' self-interaction branch, 
and an interaction with the mean flow and matches numerical simulations 
(Figures s\ref{fig:k-w_simwater} and \ref{fig:num_k_w}) and observed sea surface spectra (Figures \ref{fig:k-w_ww}). 
It confirms that, in the absence of time resonances, 
space-resonances dominate the nonlinear dynamics for intermediate times
and their signature persists for long times.
We note that the delta functions around the space resonances appearing in the approximate expression for the spatio-temporal spectrum \eqref{eq:finite_spectrum}, e.g $\delta(\sigma-2\omega(k/2))$, are a crude approximation of a complicated function of $(k,\sigma)$ that is highly peaked around space resonances 


These results pave the way for analyzing more realistic wave systems where higher-order interactions are present, and exact (time) resonances do occur there.
In particular, the water wave problem exhibits both time and space resonances and introduces additional complexity due to its rich Hamiltonian structure and nonlinearity. In Section \ref{sec:waterwaves}, we will apply the framework developed here to the water wave system and explore how these resonance mechanisms manifest in its dynamics and spectral signatures.

\subsection*{
Other three-waves interactions and negative frequencies}

Before turning to the problem of water waves, we address in more details the generation of negative $\sigma$ frequencies in the spatio-temporal spectrum. We start by remarking that the negative-frequency spectrum can in general be generated by (i) space resonance of three-wave ``summation'' interactions; (ii) three-wave interactions with zeroth mode; and (iii) space resonance of gauge-breaking four-wave interactions, all of which we will discuss in this paper. To be more specific, the negative frequencies in figure \eqref{fig:k-w_simwater} are generated from mechanisms (i) and (iii), and those in figure \eqref{fig:num_k_w} are from (ii) and (iii).

In this section, we stay in the framework of the simplified model. However, since our immediate goal is to explain the negative frequencies generated from three-wave processes, i.e., (i) and (ii), we make the model more general by augmenting it with the interaction term (in order to explain (i))
\begin{align}\label{gaugeH}
\frac{\epsilon}{3}\sum_{k_{1}k_{2}k_{3}} V^{k_{1}k_{2}k_{3}} 
  \left(a_{1}^{*}a_{2}^{*}a_{3}^{*} + a_{1}a_{2}a_{3}\right) \delta_{k_{1}k_{2}k_{3}}
\end{align}
where we again omit factors of $L^{-d}$ for light notation and $\delta_{k_{1}k_{2}k_{3}}=\delta(k_1+k_2+k_3)$.
Filtering out the linear dynamics using \eqref{eq:b_env}, the Hamiltonian \eqref{gaugeH}
contributes a term 
in the leading-order expansion \eqref{eq:b_leading} of $b_k(t)$
\begin{align}\label{bgaugeH}
b_{k}\left(t\right) & = \dots - \epsilon \sum_{k_{2},k_{3}}B^{kk_{2}k_{3}} 
  e^{i\omega^{k23}t}\delta_{kk_{2}k_{3}} + O\left(\epsilon^{2}\right),
  \qquad B^{kk_{2}k_{3}} 
  := \frac{V^{kk_2k_3}
  }{\omega_{k23}}b_{2}^{0*}b_{3}^{0*}.
\end{align}
This, in turn, contributes to the spatio-temporal Fourier amplitude $\tilde{a}_k(\sigma)$ for $\sigma$ negative:
\begin{align}\label{eq:spatio_amplitude'}
    \tilde{a}_k\left(\sigma\right) = ... - \epsilon\sum_{k_{2},k_{3}} B^{kk_{2}k_{3}} \delta_{kk_{2}k_{3}}
    \delta\left(\sigma+\omega_{2}+\omega_{3}\right) + O\left(\epsilon^{2}\right).
\end{align}
The associated space resonances occur when $k_2 = k_3 = -k/2$, yielding support on the curve 
\begin{align}\label{signeg1}
\sigma = -2\omega(-k/2). 
\end{align}
For surface gravity waves, this corresponds to $\sigma = -\sqrt{2g|k|}$, consistent with the signatures seen in 
both Figures \ref{fig:k-w_ww} and Figure \ref{fig:k-w_simwater}.
Note, however, how the signature \eqref{signeg1} appears to be much weaker in the numerics for the model \eqref{eq:3waves}
in Figure \ref{fig:num_k_w}; this is exactly because the Hamiltonian for the model \eqref{eq:3waves} does not include the term 
\eqref{gaugeH}. Nevertheless, frequencies concentrated around the curve \eqref{signeg1} are still generated at lower order
by more perturbative terms in the expansion.

\medskip
Another source of negative $\sigma$ frequencies comes from interactions with the mean flow ($k = 0$). 
For simplicity, when considering the simplified model \eqref{eq:Hsimple} above,
we have disregarded these 
interactions. 
Formally this can be done by imposing that the symbols of the interactions vanish
and, more precisely, that they satisfy (recall the notation \eqref{eq:b_leadingcoeff})
\begin{align}\label{limits}
\begin{split}
& \lim_{k_2\rightarrow 0} |B^{k}_{k_2k_3}|, \quad \lim_{k_3\rightarrow 0} |B^{k}_{k_2k_3}|,
  \quad \lim_{k \rightarrow 0} |B^{kk_3}_{k_2}|, \quad \lim_{k_3 \rightarrow 0} |B^{kk_3}_{k_2}| = 0
\end{split}
\end{align}
This is indeed what happens for the gravity waves system that we are going to consider in Section \ref{sec:waterwaves}.

In contrast, in the numerically simulated models \eqref{eq:3waves}, the coefficients $V^{k_1}_{k_2k_3}$
and $V^{k_1k_3}_{k_2}$ are non-zero constants and, therefore, the limits in \eqref{limits} diverge.
Nevertheless, the theory developed here 
provides some insights and confirmation of the numerical findings in these cases as well.
Indeed, in the presence of interactions with the zero mode in \eqref{eq:Hsimplemotion}, our theory suggests:

\setlength{\leftmargini}{1.5em}
\begin{itemize}

\item[-] An intense signature around the origin in the $(k,\sigma)$ plane on the line $\sigma = -\omega'(k_f)k$,
in view of \eqref{eq:b_estimate} and \eqref{Doppler},
and the fact that the coefficients $B^{k}_{k_2k_3}$ and $B^{k_1k_3}_{k_2}$ are very large 
close to the origin; this is in perfect agreement with Figure \ref{fig:num_k_w};

\item[-] A contribution to the spectrum at order $O(\epsilon^2)$ concentrated on the curve 
$$\sigma = -\omega(-k) + \omega(0),$$ 
albeit not due to space resonances, from the terms (see \eqref{eq:spatio_amplitude} and \eqref{eq:spatio_amplitude'} respectively)
\begin{align*}
- 2\epsilon\sum_{k_{2},k_{3}} B^{kk_3}_{k_2}
  \delta_{kk_{3}}^{k_{2}}\delta\left(\sigma-\omega_{2}+\omega_{3}\right), 
  \qquad 
  - \epsilon\sum_{k_{2},k_{3}} B^{kk_{2}k_{3}} \delta_{kk_{2}k_{3}}
    \delta\left(\sigma+\omega_{2}+\omega_{3}\right)
\end{align*}
and the non-resonant interactions $k_2 = 0$, $k_3=-k$.
Once again, in the absence or suppression of the zero mode, negative frequencies $\sigma = -\omega(-k)$ 
can still be generated but at lower order $O(\epsilon^4)$; see \eqref{sigma<0four} below.

\end{itemize}


To summarize, spectral curves with negative temporal frequencies, such as 
$\sigma = -\omega(-k)$ or $\sigma = -2\omega(-k/2)$ 
can be generated at leading perturbative order by a three-waves Hamiltonian \eqref{gaugeH},
and also by \eqref{eq:Hsimple} provided the zero mode is activated.
Strong signatures around the origin on lines with slope 
given by the group velocity of the dominant spatial frequency are also generated in the presence of a zero mode.


\medskip
We are now prepared to derive the spatio-temporal spectrum of weakly interacting water waves.


\bigskip
\section{The Water Wave Problem}\label{sec:waterwaves} 
The water wave problem is a fundamental example of a nonlinear, dispersive, 
Hamiltonian system with rich resonant structure. 
Importantly, it is exactly in the context of these equations that Hasselmann \cite{hasselmann62,hasselmann63b,hasselmann63c} 
developed a kinetic theory.
Unlike the simplified models considered in Section \ref{sec:space_res}, 
the water wave equations naturally involve higher-order nonlinearities 
which support both exact time resonances  and space resonances.


\subsection*{The Hamiltonian}
In parallel with Section \ref{sec:space_res}, 
our starting point is the Hamiltonian for gravity water waves in
$3$d (the case of a $2$d surface):
\begin{equation}\label{wwHam}
\mathcal{H}(\eta,\psi) := \frac{1}{2} \int G(\eta)\psi \cdot \psi \, dx + \frac{g}{2} \int \eta^2 \, dx,
\end{equation}
where, as in Section \ref{sec:expnum}, $\eta$ represents the surface elevation, $\psi$
is the restriction of the velocity potential to the interface, and $G(\eta)$ is the Dirichlet-Neuman operator associated to the fluid domain,
$G(\eta)\psi = - \eta_x\psi_x+(1+\eta_x^2)\phi_z$.
The integrals are over a large torus $(\T/L)^2$ and $g$ is the gravitational constant.
$(\eta,\psi)$ are canonical symplectic conjugate variables, and 
Hamilton's equations of motions associated to \eqref{wwHam} are equivalent to \eqref{eq:eta}-\eqref{eq:phi}.

As it is commonly done in the theory of water waves, we work with the variable $a_k$ as in \eqref{eq:ak},
which we recall here for ease of reference:
\begin{equation}\label{eq:ak'}
a(k,t)=\sqrt{\frac{g}{2\omega(k)}}\eta(k,t) - i \sqrt{\frac{\omega(k)}{2g}}\psi(k,t),
  \qquad \omega(k)=\omega_k=\sqrt{g|k|}.
\end{equation}
$\omega(k)$ is the linear dispersion relation and the variable \eqref{eq:ak'}
diagonalizes the quadratic part of the Hamiltonian. 
Performing a Taylor expansion of the Dirichlet-Neumann map for small amplitudes,
one can write an expansion for the Hamiltonian in terms of the variable $a_k$ as follows \cite{zakharov1999statistical}:
\begin{align}\label{Hn}
\begin{split}
\mathcal{H} & = \mathcal{H}_0(a,a^*) + \mathcal{H}^{(1,2)}(a,a^*) + \mathcal{H}^{(0,3)}(a,a^*)
  \\ & + \mathcal{H}^{(2,2)}(a,a^*) 
  + \mathcal{H}^{(1,3)}(a,a^*) + \mathcal{H}^{(0,4)}(a,a^*) + \cdots
\end{split}
\end{align}
where 
\begin{align}\label{Hn0}
\mathcal{H}_0 & = \frac{1}{2} \sum_k \omega_k a_k a_k^*, 
\\
\label{Hn12}
\mathcal{H}^{(1,2)} & := \frac{1}{2} \sum V^{k}_{k_1,k_2} 
  \big( a_{k} a_{k_1}^* a_{k_2}^* + a_{k}^* a_{k_1} a_{k_2} \big)
  \delta^k_{k_1k_2} 
\\
\label{Hn03}
\mathcal{H}^{(0,3)} & := \frac{1}{6} \sum V^{k,k_1,k_2}
  \big( a_{k} a_{k_1} a_{k_2} + a_{k}^* a_{k_1}^* a_{k_2}^* \big)
  \delta^{kk_1k_2},
\\
\label{Hn22}
\mathcal{H}^{(2,2)} & := \frac{1}{2} \sum V^{k,k_1}_{k_2,k_3}
  \big( a_{k} a_{k_1} a_{k_2}^* a_{k_3}^* + a_{k}^* a_{k_1}^* a_{k_2} a_{k_3} \big)
  \delta^{kk_1}_{k_2k_3}
\\
\label{Hn13}
\mathcal{H}^{(1,3)} & := \frac{1}{6} \sum V^{k}_{k_1,k_2,k_3} 
  \big( a_{k} a_{k_1}^* a_{k_2}^* a_{k_3}^* + a_{k}^* a_{k_1} a_{k_2} a_{k_3} \big)
  \delta^k_{k_1k_2k_3}
\\
\label{Hn04}
\mathcal{H}^{(0,4)} & := \frac{1}{8} \sum V^{k,k_1,k_2,k_3}
  \big(a_{k} a_{k_1} a_{k_2} a_{k_3} 
  +  a_{k}^* a_{k_1}^* a_{k_2}^* a_{k_3}^* \big)
  \delta^{kk_1k_2k_3}
\end{align}
and we disregard terms of order $O(|a|^5)$.
The symbols $V$ are all real-valued and assumed to be fully symmetrized without loss of generality
(e.g. $V^{k}_{k_1,k_2} = V^{k}_{k_2,k_1}$, $V^{k,k_1,k_2} = V^{\pi(k,k_1,k_2)}$ for any permutation $\pi$,
and so on).
We will sometimes abbreviate these symbols by replacing the upper or lower indexes $k_j$ just by $j$ 
when this causes no confusion. 
Let us remark a few properties of these symbols:

\setlength{\leftmargini}{1.5em}

\begin{itemize}

\item[(1)] {\it There are no interactions with the zero mode}, and the symbols vanish if any of the wave numbers is zero.

\item[(2)] Since the dispersion relation of water waves is a concave function of the wavenumber \( k \), 
{\it there are no non-trivial three-wave time resonances}, i.e. no non-trivial solutions of \eqref{timeres}. 
The trivial solutions (e.g. $k_2=0$) which, in principle, could be problematic, are 
completely suppressed by the vanishing property of the symbols in the sense that 
(cfr. with \eqref{limits} and recall the notation $V^{k2}_{1}=V_{k2}^{1}$)
\begin{align}\label{limitsww}
\begin{split}
& \lim_{k_2\rightarrow 0} \Big|\frac{V^{k}_{12}}{\omega^k_{12}}\Big|
  + \lim_{k \rightarrow 0} \Big|\frac{V^{k2}_{1}}{\omega^{k2}_1}\Big|
  + \lim_{k_2 \rightarrow 0} \Big|\frac{V^{k2}_{1}}{\omega^{k2}_1}\Big| = 0.
\end{split}
\end{align}

\item[(3)] {\it Non-trivial four-waves time resonances exists}
and the respective interaction symbols do not vanish,
also after the non-resonant 
three-waves interactions are effectively eliminated following the integration by parts procedure 
described in \eqref{eq:first_nl}-\eqref{eq:b_leading} or, equivalently, after performing 
a normal form transformation 
at the level of the Hamiltonian that eliminates $\mathcal{H}^{(1,2)}$ and $\mathcal{H}^{(0,3)}$.

\end{itemize}


\smallskip
In the remaining of this section we identify the leading-order nonlinear corrections to the spatio-temporal 
spectrum of water waves. 
These include contributions from \textit{space resonances} in both three- and four-waves interactions
which give rise to additional high-intensity curves in the spectrum,
and four-waves \textit{space-time resonances} which modify the dispersion relation beyond linear order,
giving rise to a so-called {\it phase shift}.
We anticipate that a key qualitative difference between the full Hamiltonian for water waves, \eqref{Hn}, 
and the simplified models, \eqref{eq:Hsimple}, lies in the presence of terms like \eqref{Hn03} (or \eqref{gaugeH}
as in the earlier discussion).
These terms give rise to space resonances that manifest as contributions 
at \textit{negative frequencies}, $\sigma<0$, in the spatio-temporal spectrum \cite{villois2025anomalous}. 
To our knowledge, this phenomenon has not previously been analyzed in the context of water waves.



\subsection*{Approximate evolution}
Starting from the evolution equation $\partial_t a_k = -i (\partial \mathcal{H})/(\partial a_k^*)$,
we filter the linear motion and introduce a small parameter by letting
$b(t,k) = b_{k}(t) := \epsilon^{-1}a_{k}(t)e^{i\omega_{k}t}$, 
and then write the leading order expansion for the amplitude $b_k(t)$ 
after integration by parts in time of the non-resonant quadratic terms, as done in \eqref{eq:first_nl},
and integrating by parts the non-resonant cubic terms as well.
With this procedure we can split $b_k(t)$ 
into two main contributions as
\begin{align}\label{wwsplitb}
b_k\left(t\right) = b_{k}^{L}\left(t\right) + b^{O}_k\left(t\right) 
\end{align}
where $b_{k}^{L}\left(t\right)$ ($L$ for Linear) 
has spatio-temporal spectrum supported in a neighborhood of the (shifted) linear dispersion branch $\sigma = \sqrt{g|k|} + \epsilon^2 \delta \omega_k$,
and $b_{k}^{O}\left(t\right)$ ($O$ for Outside) has support outside the linear branch.
We will determine all the corrections up to (and including) order $O(\epsilon^4)$ 
in the spatio-temporal spectrum of $b^O$.
For $b^{L}$, instead, we will only include corrections to the amplitude and phase shift up to (and including) $O(\epsilon^2)$;
we do not analyze other corrections to $b^L(t)$ that are coming from non-trivial $4$-waves time-resonances
and are expected to be driven by the corresponding wave-kinetic equation.

The profiles in \eqref{wwsplitb} are defined as follows
(cfr. the case of the simpler model \eqref{eq:b_leading}-\eqref{eq:b_leadingcoeff}):
\begin{subequations}\label{eq:b_water_leading}
\begin{align}
\label{eq:b_water_leading1}
b_{k}^O(t) & :=
  -\frac{\epsilon}{2} \sum_{k_{2},k_{3}} B^{k}_{k_{2}k_{3}} 
  e^{i\omega_{23}^{k}t}\delta_{k_{2}k_{3}}^{k}
  - \epsilon\sum_{k_{2},k_{3}} B_{k_2}^{kk_{3}} 
  e^{i\omega_{2}^{k3}t}\delta_{kk_{3}}^{k_{2}}
  - \frac{\epsilon}{2} \sum_{k_{2},k_{3}} B^{kk_{2}k_{3}} 
  e^{i\omega^{k23}t} \delta^{kk_{2}k_{3}} 
  \\
  \label{eq:b_water_leading3}
  & -
  \frac{\epsilon^{2}}{2} \sum_{k_{2},k_{3},k_{4}} B^{kk_3k_4}_{k_2} 
  e^{i\omega_{2}^{k34}t}\delta_{kk_{3}k_{4}}^{k_{2}}
  - \frac{\epsilon^{2}}{2} \sum_{k_{2},k_{3},k_{4}} B^{kk_{2}k_{3}k_{4}} 
  e^{i\omega^{k234}t} \delta^{kk_{2}k_{3}k_{4}}
  \\
  \label{eq:b_water_leading4}
  & - \frac{\epsilon^{2}}{6} \sum_{k_{2},k_{3},k_{4}} B^{k}_{k_2k_3k_4}
  e^{i\omega_{234}^{k}t}\delta_{k}^{k_{2}k_{3}k_{4}}
  + \mathcal{O}_{Wick}\left(\epsilon^{3}\right)
  + \mathcal{R}^O(k,t)
\end{align}
\end{subequations}
and
\begin{align} \label{eq:b_water_leading2}
b_{k}^{L}\left(t\right) := B_{k}^{\left(0,1\right)} + i t \epsilon^{2} 
  \Big( 2\sum_{k_{2}\neq k} M^{kk_2}_{kk_2} \left| b_{k_2} \right|^2 
  + M^{kk}_{kk} \left| b_{k} \right|^2 \Big) b_k  
  \\
  + \epsilon^2 \int_0^t \mathcal{C}\left(b(s),b(s),b^\ast(s) \right) \, ds 
  + \mathcal{R}^L(k,t) 
\end{align}
where we now define all the terms in \eqref{eq:b_water_leading}-\eqref{eq:b_water_leading2}:

\setlength{\leftmargini}{1.5em}
\begin{enumerate}

\item 
 In \eqref{eq:b_water_leading} we are  using the same notation from \eqref{eq:b_leadingcoeff} 
 and \eqref{bgaugeH} for the cubic symbols in \eqref{eq:b_water_leading1}
 and adopt a similar notation (upper indexes correspond to complex conjugates)
 for the quartic symbols, that is, we let
\begin{align}\label{wwB}
\begin{split}
B^{kk_{3}k_{4}}_{k_2} 
  & := \frac{1}{\omega^{k34}_{2}} \Big( V^{kk_3k_4}_{k_2} + \cdots \Big) b_{2}^{0}b_{3}^{0*}b_{4}^{0*},
\\
B^{k}_{k_2k_3k_4} 
  & :=  \frac{1}{\omega^{k}_{234}}\Big( V^{k}_{k_2k_3k_4} + \cdots \Big) b_{2}^{0}b_{3}^{0}b_{4}^{0},
\\
B^{kk_{2}k_{3}k_{4}} 
  & := \frac{1}{\omega^{k234}} \Big( V^{kk_2k_3k_4} + \cdots \Big) b_{2}^{0*}b_{3}^{0*}b_{4}^{0*},
\end{split}
\end{align}
 where ``$\cdots$'' denote other expressions
 that can be determined only from the three-waves coefficients 
 and that arise from the normal form transformations of the cubic terms;
 for example (conjugates on the symbols are included just for bookkeeping here)
\begin{align}
\begin{split}
B^{kk_3k_4}_{k_2} := \frac{1}{\omega^{k34}_{2}} \left[ 
    \frac{1}{2}V^{kk_3k_4}_{k_2}
 + \frac{1}{2} \sum_{k_5} \Big( \frac{V^{kk_5}_{k_2}}{\omega^{k5}_{2}} \overline{ V^{k_5}_{k_3k_4} }
 + 2\frac{V^{kk_3k_5}}{\omega^{k35}} \overline{V^{k_5k_2}_{k_4}}
 + \frac{V^{kk_3}_{k_5}}{\omega^{k3}_{5}} V^{k_5k_4}_{k_2}
 + \frac{V^{k}_{k_2k_5}}{\omega^{k}_{25}} V^{k_5k_3k_4} \Big)
 \right] \\ \times b_{2}^{0}b_{3}^{0*}b_{4}^{0*}.
\end{split}
\end{align}
Note that all denominators in \eqref{wwB} are non-zero by the concavity of $\omega_k$.
The exact form of the coefficients in \eqref{wwB} is unimportant for our analysis.

\item The $\mathcal{O}_{Wick}(\epsilon^{3})$ term in \eqref{eq:b_water_leading4} denotes 
quartic terms in the amplitude's profile $b$ that drop out in our computation of the spectrum
under Wick contraction.
Indeed, since these terms are given by a coefficient times four independent gaussians, 
their only contribution at order $O(\epsilon^4)$ in the spectrum can 
come from pairing them with the quadratic terms from \eqref{eq:b_water_leading1}
(two independent gaussians); however, such a pairing cannot happen because of the lack of 
non-trivial three waves resonance and the absence of the zero mode. 

\item The $\mathcal{R}^O$ term in \eqref{eq:b_water_leading4} denotes $O(\epsilon^{5})$
terms in the perturbative expansion whose contributions we disregard since, for $\epsilon$ small enough,
they are barely visible in the spectrum.

\item In \eqref{eq:b_water_leading2} we define
\begin{align}\label{wwB01}
\begin{split}
B_{k}^{\left(0,1\right)} & := b_k^0  
  + \frac{\epsilon}{2}\sum_{k_{2},k_{3}} B^{k}_{k_{2}k_{3}} 
    \delta_{k_{2}k_{3}}^{k} + \epsilon \sum_{k_{2},k_{3}}
    B^{kk_{3}}_{k_{2}} 
    \delta_{kk_{3}}^{k_{2}}
    + \frac{\epsilon}{2}\sum_{k_{2},k_{3}}
    B^{kk_{2}k_{3}} 
    \delta^{kk_{2}k_{3}} 
    \\ 
    & 
  + \frac{1}{2}\epsilon^{2} \sum_{k_{2},k_{3},k_{4}} B^{kk_3k_4}_{k_2} 
  \delta_{kk_{3}k_{4}}^{k_{2}}
  + \frac{1}{2}\epsilon^{2} \sum_{k_{2},k_{3},k_{4}} B^{kk_{2}k_{3}k_{4}} 
 \delta^{kk_{2}k_{3}k_{4}}
  \\
  & + \frac{1}{6}\epsilon^{2} \sum_{k_{2},k_{3},k_{4}} B^{k}_{k_2k_3k_4}
  \delta_{k}^{k_{2}k_{3}k_{4}}
  \\ 
\end{split}
\end{align}
where we are adopting the same notation for the symbols;
this is the analogue \eqref{eq:b_leadingcoeff0} for the simpler model 
but we are now accounting for the additional three- and four-waves interactions. 

\item 
The second term on the right-hand side of \eqref{eq:b_water_leading2} is a `phase shift' contribution
coming from trivial resonances for the `gauge-invariant' cubic terms. 
The value of the symbol can be calculated from
\begin{align}\label{Mgauge}
\begin{split}
M^{kk_4}_{k_2k_3} := 
\frac{1}{2}V^{kk_4}_{k_2k_3}
 + \frac{1}{2} \sum_{k_5} \Big( \frac{V^{kk_4}_{k_5}}{\omega^{k4}_{5}} V^{k_5}_{k_2k_3} 
 + 2\frac{V^{kk_5}_{k_2}}{\omega^{k5}_{2}} \overline{V^{k_5k_3}_{k_4}}
 + 2 \frac{V^{k}_{k_2k_5}}{\omega^{k}_{25}} V^{k_5k_4}_{k_3}
 + \frac{V^{kk_2k_5}}{\omega^{k25}} \overline{V^{k_5k_3k_4}}\Big).
\end{split}
\end{align}
As we will describe below, this term contributes an $O(\epsilon^2)$ 
shift to the main spectral curve $\sigma = \omega(k)$;
see also the discussion at the end of Section \ref{sec:space_res} in the context of the simpler model.

\item The time integral in \eqref{eq:b_water_leading2}
contains all cubic interactions that are (globally) gauge invariant, i.e., trilinear expressions in $b,b,b^\ast$; 
these are explicitly given
\begin{align}
\mathcal{C}\left(b(s),b(s),b^\ast(s) \right) := \sum_{k_2,k_3 \neq k_4} M^{kk_4}_{k_2k_3} \delta^{kk_{4}}_{k_2k_{3}} 
 \, b_{k_2}(s)b_{k_3}(s)b_{k_4}^*(s)
\end{align}
with the definition of the symbol in \eqref{Mgauge}.
These terms give contributions 
to the amplitude that are expected to be (mainly) supported on the linear branch;
they are related to weak wave turbulence theory and are the subject 
of ongoing investigation.

\item Finally, $\mathcal{R}^L(k,t)$ includes nonlinear expressions that are 
of quintic or higher homogeneity in $(b,b^*)$, that appear with an $\epsilon^4$ or higher power in front,
and that contribute (mostly) to the linear branch of the spatio-temporal spectrum.

  
\end{enumerate}


\subsection*{Space resonances and signatures in the spatio-temporal spectrum}
We now analyze the spatio-temporal spectrum of $b_k^O(t)$.

\medskip
\noindent
{\bf 3-Waves.}
The analysis of the three-waves interactions can be carried out exactly as described
in Section \ref{sec:space_res}, taking into account the presence of all the 3-waves terms as in
\eqref{eq:Hsimple} and \eqref{gaugeH}, and the absence of the zero mode.
This yields, at leading order beyond the linear dispersion curve, the 
curves 
\begin{align}\label{sigma<0three}
\sigma = 2\omega\left(k/2\right) = \sqrt{2g|k|}, \qquad \sigma=-2\omega\left(-k/2\right) = -\sqrt{2g|k|},
\end{align}
corresponding to the space-resonances
\begin{align*}
\nabla_{k_{2}}\left(\omega^{k}_{23}\right) \delta^{k}_{k_{2}k_{3}} & = 0 \quad \iff \quad k_{2}=k_{3}=k/2,
\\
\nabla_{k_{2}}\left(\omega_{k23}\right) \delta^{kk_{2}k_{3}} & = 0 \quad \iff \quad k_{2}=k_{3}=-k/2.
\end{align*}
This agrees with the measurements in Figure \ref{fig:k-w_ww} and the numerics in Figure \ref{fig:k-w_simwater}.
Note how the curves in \eqref{sigma<0three} 
match a ``second-harmonic'' type curve $\sigma = \pm \omega(2k)$
in the case of the specific deep gravity waves dispersion. However, 
as we showed, the emergence of these curves is not due to a Stokes expansion but actually linked to space resonances; as additional supporting evidence, in the case of the Schr\"odinger equation, $\omega = |k|^2$,  the secondary curve is $\sigma = |k|^2/2 = 2\omega(k/2)$.

\medskip
\noindent
{\bf 4-Waves.}
Let us concentrate first on the four-waves terms that break the global gauge (or phase rotation symmetry),
that is, those in \eqref{eq:b_water_leading3} and  \eqref{eq:b_water_leading4}.
For these three terms we find the space-resonances
\begin{align}
\nabla_{k_{2},k_{3}} \left(\omega^{k34}_{2}\right) \delta^{kk_{3}k_{4}}_{k_{2}} & = 0 
  \quad \iff \quad k_{2}=k_{3}=k_{4}=-k
\\
\nabla_{k_{2},k_{3}} \left(\omega^{2k34} \right) \delta^{kk_{2}k_{3}k_{4}} & = 0 
  \quad \iff \quad k_{2}=k_{3}=k_{4}=-k/3
\\
\nabla_{k_{2},k_{3}} \left(\omega^{k}_{234} \right) \delta^{k}_{k_{2}k_{3}k_{4}} & = 0 
  \quad \iff \quad k_{2}=k_{3}=k_{4}= k/3
\end{align}
which give, respectively, the contributions to the spatio-temporal spectrum 
\begin{align}\label{sigma<0four}
\sigma=-\omega(k) \quad \mbox{and} \quad \sigma=-3\omega\left(-k/3\right) \quad \mbox{and} \quad \sigma = 3\omega(k/3);
\end{align}
This explain theoretically the linear dispersion curves visible in Figures \ref{fig:k-w_ww} and \ref{fig:k-w_simwater} 
when $\sigma < 0$, and the appearance of ``third-harmonic'' curve
$\sigma = \pm \sqrt{3g|k|}$ which is also visible in Fig. \ref{fig:k-w_simwater}.

Finally, we look at the global gauge invariant interaction and see that 
\begin{align}\label{wwgaugeSR}
\nabla_{k_{2},k_{3}}\left(\omega^{kk_2}_{k_3k_4}\right) \delta_{k_{3}k_{4}}^{kk_{2}} = 0
 \quad \iff \quad k_{2}=k_{3}=k_{4}=k.
\end{align}
These space-resonant interactions are also time-resonant,
and are a subset of the `trivial' resonances 
\begin{align}\label{trivialres}
k_{3}=k \quad \mbox{and} \quad k_{4}=k_2, \qquad \mbox{or} \qquad 
k_{4}=k \quad \mbox{and} \quad k_{3}=k_2.
\end{align}
As it turns out, \eqref{trivialres} are responsible for the frequency shift. 

\subsection*{The spatio-temporal spectrum of water waves}
Putting the above observations all together, 
analogously to approximation of the spatio-temporal spectrum for the simplified model
derived in \eqref{eq:finite_spectrum} (recall also \eqref{Nk} and \eqref{eq:phase_shift}), 
we can write now the leading order spatio-temporal spectrum approximation for surface gravity waves.
With $n_k$ and $S_k$ defined analogously to \eqref{nk} and \eqref{Sk}, we have
and $S_k$ defined analogously to \eqref{nk} and \eqref{Sk}, we have
\begin{align}\label{eq:finite_spectrum_water_waves}
S_{k} & \approx \mathcal{N}_{k} \delta \left(\sigma-\omega_{k}+\epsilon^{2}\delta\omega_{k}\right)
\\
& + \frac{\epsilon^2}{4}\left\langle \left| B^{k}_{k/2,k/2} 
  \right|^{2}\right\rangle \delta\left(\sigma - 2\omega_{k/2}\right)
  + \epsilon^2 \left\langle \left| B^{k,k_f-k}_{k_f} 
  \right|^{2}\right\rangle \delta\left(\sigma-\omega'\left(k_{f}\right)k\right)
\\
& + \frac{\epsilon^2}{4}\left\langle \left|B^{k,-k/2,-k/2} 
  \right|^{2}\right\rangle \delta\left(\sigma+2\omega_{-k/2}\right)
\\ 
&  + \frac{\epsilon^4}{4}\left\langle \left|B^{k,-k,-k}_{-k} 
\right|^{2}\right\rangle \delta\left(\sigma+\omega_{-k}\right)
 + \frac{\epsilon^4}{4}\left\langle \left|B^{k,-k/3,-k/3,-k/3} 
  \right|^{2}\right\rangle 
  \delta\left(\sigma+3\omega_{-k/3}\right)
  \\
  & + \frac{\epsilon^4}{36}\left\langle \left|B^{k}_{k/3,k/3,k/3} 
  \right|^{2}\right\rangle 
  \delta\left(\sigma - 3\omega_{k/3}\right).
\end{align}
where:

\begin{itemize}

\item [-] The corrected amplitude of the linear spectrum $n_k$ in the spatio-temporal spectrum, 
given by (see \eqref{wwB01}) 
\begin{align}\label{Nkww}
\begin{split}
 \mathcal{N}_{k} := \left\langle B_{k}^{\left(0,1\right)}B_{k}^{*\left(0,1\right)}\right\rangle = n_{k} 
  & + \frac{\epsilon^2}{4}\sum_{k_{2},k_{3}}\left\langle \left| B^{k}_{k_{2},k_{3}} 
  \right|^{2}\right\rangle \delta_{kk_{3}}^{k_{2}}
  + \epsilon^2 \sum_{k_{2},k_{3}} \left\langle \left| B^{k,k_3}_{k_2} 
  \right|^{2}\right\rangle \delta_{kk_{3}}^{k_{2}}
  \\
  & +\frac{\epsilon^2}{4}\sum_{k_{2},k_{3}}\left\langle \left| B^{k,k_{2},k_{3}} 
  \right|^{2}\right\rangle \delta^{kk_{2}k_{3}} 
  + \mathcal{O}\left(\epsilon^{4}\right),
\end{split}
\end{align}

where here
\begin{align}
\mathcal{O}\left(\epsilon^{4}\right) & = \frac{\epsilon^4}{4}\sum_{k_{2},k_{3},k_{4}}\left\langle \left|B^{k,k_{2},k_{3},k_{4}} 
  \right|^{2}\right\rangle \delta^{kk_{2}k_{3}k_{4}}
  + \frac{\epsilon^4}{4} \sum_{k_{2},k_{3},k_{4}} \left\langle \left|B^{k,k_{3},k_{4}}_{k_{2}} 
  \right|^{2}\right\rangle \delta^{kk_{3}k_{4}}_{k_{2}}
  \\
  & + \frac{\epsilon^4}{36}\sum_{k_{2},k_{3},k_{4}} \left\langle \left|B^{k}_{k_{2},k_{3},k_{4}}
  \right|^{2}\right\rangle \delta^{k}_{k_{2}k_{3}k_{4}}.
\end{align}
The amplitude that multiplies the $\delta$ supported on the main branch of the spatio-temporal spectrum, which we denote by $\mathcal{N}_k$ here, is a correction to $n_k$, see (18); however, notice that this is not the same correction to the spectral second moment computed in [3, 33]. The latter can be computed using the expansion of the spectral amplitude $b^L_K(t)$ in \eqref{eq:b_water_leading2}.
The fact that there are no other $\mathcal{O}(\epsilon^2)$ contributions in \eqref{Nkww} is 
due to the following: 
(1) the suppression of trivial resonances, see \eqref{limitsww}, which means that $\mathcal{O}(\epsilon)$
terms in \eqref{eq:b_water_leading1} can only pair with their conjugates, giving rise to 
the $\mathcal{O}(\epsilon^2)$ terms in \eqref{Nkww}, and (2) that the pairings of the $\mathcal{O}(\epsilon^2)$ terms 
in \eqref{wwB01} with the linear term $(b_k^0)^*$ all vanish under Wick contraction.
Also notice that in \eqref{Nkww} we have left the dependence of the expressions on $b_k(0)$ to highlight the different orders in $\epsilon$, but one can eventually express everything in terms of the random initial conditions $a_k(0) = \epsilon b_k(0)$.


\item[-] The frequency shift up to order $\mathcal{O}(\epsilon^2)$ is given by 
\begin{align}
    \delta\omega_{k}& = 
  \Big( 2\sum_{k_{2}\neq k} M^{kk_2}_{kk_2} \left| b^{0}_{k_2} \right|^2 
  + M^{kk}_{kk} \left| b^{0}_{k} \right|^2 \Big)  
\end{align}
cfr. with the case of the simpler model \eqref{bshift} and the calculation
starting in \eqref{eq:a_freq} and leading to \eqref{eq:phase_shift}. Leading corrections to the frequency shift of water waves were also calculated in \cite{janssen2007intermediate, tick1959non, komen1980nonlinear}.

\end{itemize}

\section{Conclusion}\label{SecCon}
In this work we developed a theoretical and computational framework to interpret the spatio-temporal spectrum of weakly nonlinear dispersive waves in regimes dominated by non-resonant 
interactions. Using stereoscopic measurements of surface gravity waves, numerical simulations of the water wave equations, and simplified models, we identified spectral branches that cannot be explained by classical time resonances alone.
Our analysis demonstrates that space resonances, interactions between wave packets sharing the same group velocity, provide a unifying explanation for the observed features of the spectrum. In particular, they account for the appearance of higher-order branches, linear signatures near the zero mode, and negative-frequency components in systems where gauge invariance is broken. Moreover, in the absence of exact three-wave time resonances, space resonances govern the leading-order nonlinear corrections to the spectrum.
By applying this framework to the full water wave problem, we showed how three- and four-wave interactions combine to generate the dominant spectral excitations, including negative-frequency branches and ``higher harmonics", in agreement with both experimental and numerical data.
Taken together, these results establish space resonances as a central mechanism shaping the long-time dynamics and spectral signatures of nonlinear waves. They complement the classical wave turbulence picture based on time resonances and open the way to a more complete statistical description of nonlinear wave systems, with direct applications to laboratory and oceanic observations.

\begin{acknowledgments}
This research was supported by Simons Collaboration on
Wave Turbulence, Grant No. 617006.
M.O. is also supported by INFN (MMNLP and FieldTurb). 
We thank A. Benetazzo for preliminary discussions on    the data from the Acqua Alta oceanographic tower. We thank Gregory Falkovich and Oliver B\"uhler for helpful discussions.
F.P. was supported in part by NSERC grants RGPIN-2018-06487 and  RGPIN-2025-06419

\end{acknowledgments}

\bibliography{internal}

\newcommand{\brak}[1]{\langle #1\rangle}
\newcommand{\wh}{\widehat}


\end{document}